\documentclass[preprint2]{aastex}

\newcommand{\mbh}{$M_{\rm BH}$}
\newcommand{\ms}{$M_{\rm sph}$}
\newcommand{\ls}{$L_{\rm sph}$}
\newcommand{\lh}{$L_{\rm host}$}

\begin{document}

\title{COSMIC EVOLUTION OF BLACK HOLES AND SPHEROIDS. IV. THE
\mbh~- $L_{\rm sph}$ RELATION}

\shorttitle{Evolution of the \mbh~- $L_{\rm sph$} Relation}
\shortauthors{Bennert et al.}

\author{Vardha Nicola Bennert\altaffilmark{1},
Tommaso Treu\altaffilmark{1,2}, Jong-Hak Woo\altaffilmark{3,4,5}, 
Matthew A. Malkan\altaffilmark{3}, Alexandre Le Bris\altaffilmark{1,6},
Matthew W. Auger\altaffilmark{1}, Sarah Gallagher\altaffilmark{7}, 
Roger D. Blandford\altaffilmark{8}}

\altaffiltext{1}{Department of Physics, University of California, Santa
Barbara, CA 93106; bennert@physics.ucsb.edu, tt@physics.ucsb.edu, mauger@physics.ucsb.edu}

\altaffiltext{2}{Sloan Fellow, Packard Fellow}

\altaffiltext{3}{Department of Physics and Astronomy, University of California,
Los Angeles, CA 90095; woo@astro.ucla.edu, malkan@astro.ucla.edu}

\altaffiltext{4}{Hubble Fellow}

\altaffiltext{5}{Department of Astronomy, Seoul National University, Korea}

\altaffiltext{6}{Universit{\'e} Paul Sabatier, Toulouse, France; alexandrelebris@gmail.com}

\altaffiltext{7}{Department of Physics and Astronomy, University of Western
Ontario, 1151 Richmond St, London, ON N6A 3K7, Canada; sgalla4@uwo.ca}

\altaffiltext{8}{Kavli Institute for Particle Astrophysics and Cosmology,
Stanford, CA 94305; rdb@slac.stanford.edu}

\shortauthors{Bennert et al.}

\begin{abstract}
From high-resolution images of 23 Seyfert-1 galaxies at z=0.36 and
z=0.57 obtained with the Near Infrared Camera and Multi-Object
Spectrometer on board the {\it Hubble Space Telescope} (HST), 
we determine host-galaxy morphology, nuclear luminosity,
total host-galaxy luminosity and spheroid luminosity. 
Keck spectroscopy is used to estimate black
hole mass (\mbh). We study the cosmic evolution of the 
\mbh-spheroid luminosity (\ls) relation.
In combination with our previous work, totaling 40 Seyfert-1 galaxies, 
the covered range in BH mass is substantially increased,
allowing us to determine for the first time intrinsic
scatter and correct evolutionary trends for selection
effects.  We re-analyze archival HST images of 19 local reverberation-mapped active galaxies
to match the procedure adopted at intermediate redshift.
Correcting spheroid luminosity for passive luminosity evolution and
taking into account selection effects, we determine that at fixed
present-day V-band spheroid luminosity, \mbh/\ls $\propto$ $(1+z)^{2.8 \pm
1.2}$. When including a sample of 44 quasars out to $z=4.5$ taken from the literature,
with luminosity and BH mass corrected to a self-consistent calibration, 
we extend the BH mass range to over two orders of magnitude,
resulting in \mbh/\ls $\propto$ $(1+z)^{1.4 \pm 0.2}$.
The intrinsic scatter of the relation, assumed constant with
redshift, is 0.3$\pm$0.1 dex ($<$0.6 dex at 95\% CL). 
The evolutionary trend suggests that BH growth
precedes spheroid assembly. Interestingly, 
the \mbh-{\it total host-galaxy} luminosity relation is apparently non-evolving.
It hints at either a more fundamental relation or that the spheroid grows by a redistribution of stars.
However, the high-z sample does not follow this relation, indicating 
that major mergers may play the dominant role in growing spheroids above $z \simeq 1$.
\end{abstract}


\section{INTRODUCTION}
\label{sec:intro}

Supermassive black holes (BHs) seem to be ubiquitous in the center of
spheroids -- elliptical galaxies and classical bulges of spirals
\citep[e.g.,][]{kor95,fer05}.
In the local Universe, tight empirical relations have been found
between the mass of the BH (\mbh) and the properties of the
spheroid, i.e.~stellar velocity dispersion $\sigma$
\citep{fer00,geb00}, stellar mass \citep[e.g.,][]{mar03}, 
and luminosity \citep[e.g.,][]{har04}.  The tightness of these
relations is surprising, given the very different scales involved --
from accretion onto the BH ($\mu$pc scale), the dynamical sphere of
influence of the BH (pc scale) to the size of the spheroid (kpc scale)
-- and poses a challenge to any theoretical model explaining their
origin.  In general, the correlations are believed to indicate a close
connection between galaxy formation and evolution and the growth of
the BH. A variety of theoretical models have been developed to
explain the observed relations, involving galaxy mergers and nuclear
feedback through quenching of star formation
\citep[e.g.,][]{kau00,vol03,cio07,hop07,dim08,hop09b}.  

Measuring the evolution with redshift of these correlations
constrains theoretical interpretations and provides important
insights into their origin \citep[e.g.,][]{cro06,rob06,hop07}. For
quiescent galaxies, the biggest challenge is to measure the BH mass,
given the pc-scale sphere of influence of the BH which needs to be
resolved spatially through either gas or stellar dynamics
(see \citealt{gue09} and \citealt{gra08} for a recent compilation and references
therein; for a review see \citealt{fer05} and references therein)
 or from X-ray
spectroscopy probing the existence of a central temperature
peak of the interstellar medium \citep[e.g.,][]{bri99,hum06,hum08}.
With current technology, direct quiescent
black hole mass measurements are thus limited to nearby galaxies.

For active galaxies, for which nuclear luminosity is comparable to
or larger than that of the host galaxy, the situation is virtually
the opposite.  Estimating BH masses within a factor of 2-3 is fairly
straightforward through empirically calibrated relations based on
spectroscopic data measuring the kinematics of the broad-line region (BLR)
\citep[e.g.,][]{wan99,woo02,ves02,ves06,mcg08}.
Unfortunately, the active galactic nuclei (AGN) 
often outshines the host galaxy, making it
difficult to disentangle nuclear and host-galaxy light for an accurate
measurement of the spheroid luminosity. Also, measuring $\sigma$ from
stellar absorption lines is hampered by the contaminating AGN
continuum and emission lines. 
Different groups have tackled these problems in distinct ways, e.g.~by
using the [\ion{O}{3}] emission line width as surrogate of $\sigma$
\citep[e.g.,][]{shi03},
or by using gravitational lensing to
super-resolve the host galaxies of quasars
\citep[e.g.,][]{pen06a,pen06b}.  Our group
\citep{tre04,woo06,tre07,woo08} has focused on Seyfert-1 galaxies 
- for which the nucleus is not as bright as for quasars -
 at moderate redshifts ($z=0.36$ and $z=0.57$, corresponding to look-back
times of $\sim$4-6 Gyrs). The non-negligible stellar light produces strong enough
absorption lines to measure $\sigma$ from unresolved spectra, as shown
by \citet{tre04} and \citet[][hereafter Paper I \& III]{woo06,
woo08}. At the same time, high resolution {\it Hubble Space Telescope} (HST) imaging allows for an
accurate determination of the AGN luminosity (for an unbiased estimate
of nuclear luminosity and hence \mbh) and spheroid luminosity (to
create the \mbh-\ls~relation; \citealt[][hereafter Paper
II]{tre07}).  We are thus able to simultaneously study both the
\mbh-$\sigma$ and \mbh-\ls~relations, allowing us 
to distinguish mechanisms causing evolution in $\sigma$ (e.g.,
dissipational merger events) and \ls~(e.g.~through
passive evolution due to aging of the stellar population,
or dissipationless mergers).

Results presented in Paper I, II, and III suggest an offset with
respect to the local relationships, which cannot be accounted for by
known systematic uncertainties. 
At a given \mbh, in the range
10$^8$-10$^9$ M$_{\odot}$, spheroids had smaller velocity dispersion
and spheroid mass 6 Gyrs ago ($z\sim0.57)$, consistent with recent
growth and evolution of intermediate-mass spheroids. Paper II
concludes that the distant spheroids have to grow by $\sim$60\% in
stellar mass ($\Delta \log M_{\rm sph}$ = 0.20 $\pm$ 0.14) at fixed
black hole mass in the next 4 billion years to obey the local
scaling relations if no significant BH growth is assumed, consistent
with the relatively low Eddington ratios.  Indeed, the HST images
reveal a large fraction of merging or interacting systems, suggesting
that gas rich mergers will be responsible for the spheroid growth.

Although tantalizing, the results presented in our previous papers
suffer from several limitations. Samples were small, and the local
comparison sample of Seyferts measured in a self-consistent manner was
even smaller than the distant sample, thus contributing substantially
to the overall error budget. The limited range in black hole
mass was insufficient to determine independently the offset of the
scaling relation and its scatter, while taking into account selection
effects. If the \mbh-$\sigma$ and \mbh-\ls~relations of active
galaxies were not as tight as those for quiescent ones, selection
effects could be mimicking evolutionary trends
\citep{tre07,lau07,pen07}.

To overcome these limitations, we have now doubled the sample size
(from 20 in Paper II to 40 total here) and expanded the covered range
of BH masses to lower masses (from $\log$ \mbh/$M_{\odot} = 8-8.8$ in
Paper II to $\log$ \mbh/$M_{\odot} = 7.5-8.8$ here). We focus on the
resulting BH mass - spheroid luminosity relation.  
The BH mass - $\sigma$ relation will be presented in a separate
paper (Woo et al.\ 2009, in preparation).  We also analyze archival
HST images of the sample of local Seyferts with reverberation-mapped
(RM) \mbh~in the same way as our intermediate-z objects, to eliminate possible systematic
offsets. Finally, we combine our results with
data compiled from the literature and treated in a self-consistent
manner to extend the redshift range over which we study evolution.
For conciseness the three samples will be referred to as
``intermediate-redshift'' sample, ``local'' sample, 
and ``high-redshift'' sample, respectively.

The paper is organized as follows. We summarize the properties of our
intermediate-redshift Seyfert sample, observations, data reduction,
and analysis in \S~\ref{sec:sample},~\ref{sec:obs},
and~\ref{sec:surface}.  
\S~\ref{sec:quan} summarizes the derived quantities,
including the derivation of \mbh~from Keck spectra. 
In \S~\ref{sec:comp}, we describe the local comparison sample consisting
of reverberation-mapped AGNs, re-analyzed here, as well as 
the high-redshift comparison sample taken from
the literature, calibrated for consistency with the other samples.  
We present our results in
\S~\ref{sec:res}, including host-galaxy morphology and
merger rates, the evolution of the \mbh-\ls~relation,
a full discussion and treatment of selection 
effects, and a relation between BH mass and host-galaxy luminosity. 
We discuss the possible implications of our findings for the origin
and evolution of the BH mass scaling relation in \S~\ref{sec:disc}.  A
summary is given in \S~\ref{sec:sum}. 
In Appendix~\ref{sec:mc}, we describe Monte
Carlo simulations used to probe our analysis and determine errors.
Appendix~\ref{sec:sersic} discusses 
the choice of the S{\'e}rsic index in the adapted 2D
surface-brightness fitting procedure.
Details on the re-analysis of the HST images of the local
RM AGNs are given in Appendix~\ref{sec:rm}.
 Throughout the paper, we assume
a Hubble constant of $H_0$ = 70\,km\,s$^{-1}$\,Mpc$^{-1}$,
$\Omega_{\Lambda}$ = 0.7 and $\Omega_{\rm M}$ = 0.3. Magnitudes are
given in the AB system \citep{oke74}.

\section{SAMPLE SELECTION}
\label{sec:sample}

The selection of the sample of intermediate-redshift
Seyfert-1 galaxies is similar to the
one in Paper I, II, and III, with the goal to extend the sample to (a)
lower BH masses at $z \simeq 0.36$ and (b) higher redshifts of $z
\simeq 0.57$. We here briefly summarize the procedure. All objects
were selected from the Sloan Digital Sky Survey Data Release 7 (SDSS
DR7) archive according to the following criteria: (1) redshift in
either the 0.35 $<$ $z$ $<$ 0.37 bin or the 0.56 $<$ $z$
$<$ 0.58 bin, (2) H$\beta$ equivalent width and Gaussian
width $>$ 5\AA~in the rest frame.  Objects with $z \simeq 0.36$
were selected to extend the BH mass scaling relations presented in
Paper I, II, and III to the low-mass range. They meet the additional
criterion (3) \mbh~$\lesssim 10^8 M_{\odot}$ as determined from the width of
the H$\beta$ line and the $\lambda L_{\rm 5100}$ luminosity measured from the
SDSS spectra and assuming the calibration given by \citet{mcg08}.

For two objects in
Paper II (0107 and 1015) the ACS images revealed dust lanes. Thus,
Near Infrared Camera and Multi-Object
Spectrometer (NICMOS) images were additionally obtained to correct for extinction.
Table~\ref{data} summarizes the sample properties of all 23 objects.

\section{OBSERVATIONS AND DATA REDUCTION}
\label{sec:obs}

The sample was observed using the NIC2 camera and the broad filter
F110W ($\sim$ $J$-band) of NICMOS on board HST.  The 17 objects at $z$ = 0.36 were
observed for a total exposure time of 2560 seconds per object (11 as
part of GO 11208, PI Woo; 6 as part of GO 11341, PI Gallagher); the 6
objects at $z$ = 0.57 were observed for a total exposure time of 5120
seconds per object (GO 11208; PI Treu).  Four separate exposures were
obtained per object, dithering by semi-integer pixel offsets to
recover resolution lost to under-sampling and to improve cosmic-ray
and defect removal.

\begin{figure*}[h!]
\includegraphics[scale=0.9]{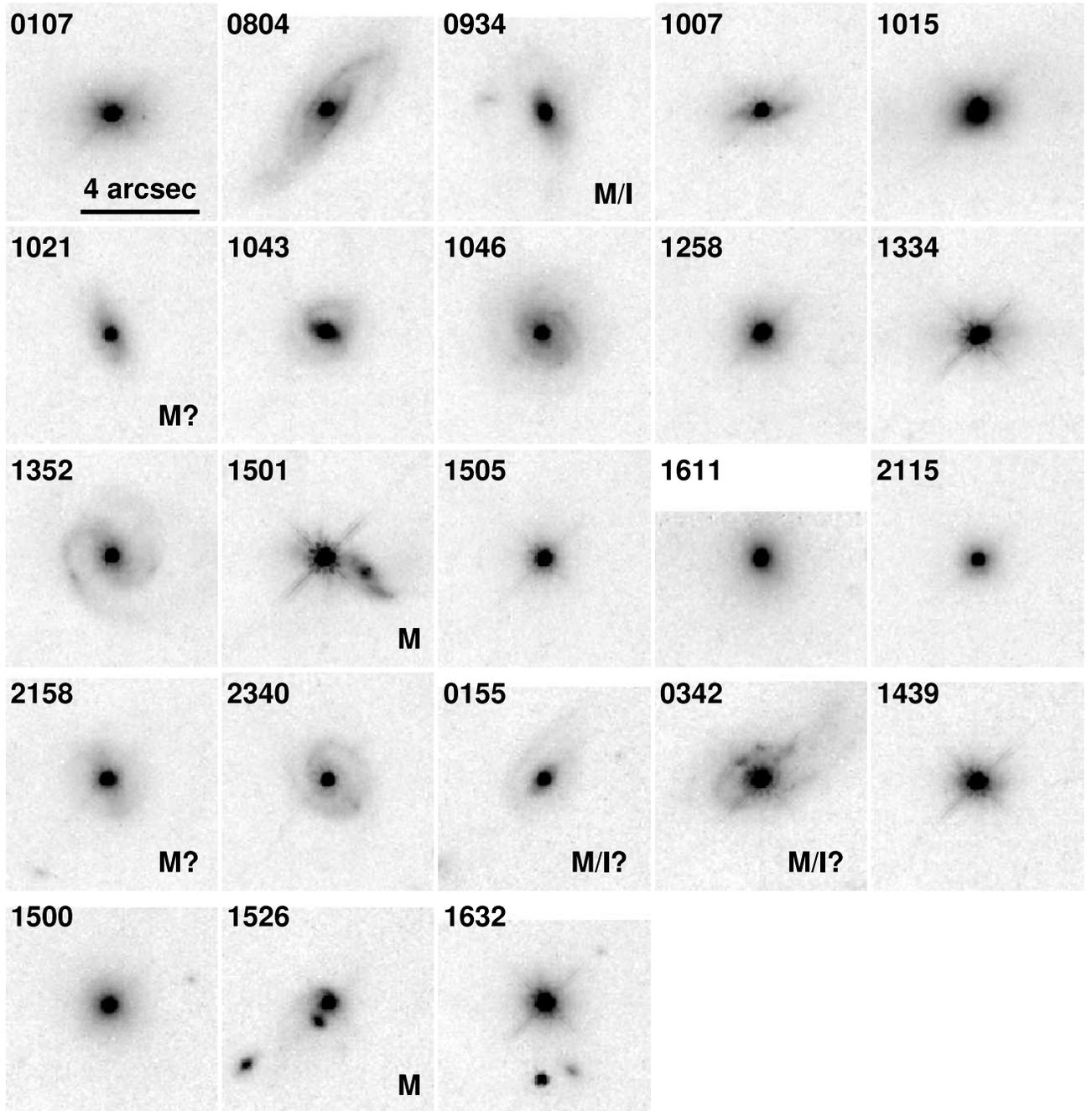}
\caption{Postage stamp NICMOS images of the 23 Seyfert-1 galaxies in the
sample. The first 17 objects are at $z$ = 0.36,
the last six objects are at $z$ = 0.57.
A 4 arcsecond scalebar is shown in the upper left image,
corresponding to $\sim$ 20 kpc at
 $z$=0.36 and 26 kpc at $z$=0.57, respectively.
The label M or M/I marks objects that are apparently merging or interacting.}
\label{nicmos}
\end{figure*}

The individual exposures were first processed with the NICMOS CALNICA
pipeline (Version 4.4.0), and then dither-combined using a custom-made
pipeline written in IRAF\footnote{IRAF (Image Reduction and Analysis
Facility) is distributed by the National Optical Astronomy
Observatories, which are operated by AURA, Inc., under cooperative
agreement with the National Science Foundation.}/STSDAS dither package
(Version 3.4.2).  The pipeline relies on the package {\sc drizzle} and
takes care of aligning the images, removing sky background, correcting
for the pedestal effect (the variable quadrant bias present in NICMOS
images) using ``pedsky''\footnote{Note that our fields are quite empty
and enough blank sky is available for an accurate determination of sky
and pedestal.}, and the NICMOS non-linearity using ``rnlincor'',
identifying and removing cosmic rays and defect pixels and finally
drizzling all input images together. For our final drizzle iteration
we chose a drizzle.pixfrac parameter of 0.9 and a drizzle.scale
parameter of 0.5, which resulted in a final scale of 0.038
arcsec/pixel.  In Figure~\ref{nicmos}, postage stamp images of all 23
objects are shown.  (We refer the reader to Paper II for the ACS
images of those 2 objects in our sample, 0107 and 1015, for which we
have both ACS and NICMOS images.)

\section{SURFACE PHOTOMETRY}
\label{sec:surface}

To decompose nuclear and host-galaxy light (spheroid and potentially
bar or disk), we used GALFIT, a 2D galaxy fitting program that can
simultaneously fit one or more objects in an image choosing from a
library of functional forms \citep[e.g.,][exponential, etc.]{ser68,
dev48} \citep{pen02}. Decomposition of complex images in multiple
components is a difficult statistical challenge due the degeneracies
involved, and the highly non-linear dependency of the likelihood on a
large number of parameters. To deal with this problem, we develop a
methodology based on physical assumptions to reduce the number of free
parameters and extensive trial and error exploration of the
multidimensional space to avoid local minima of the posterior
probability. This section describes our fitting procedure in detail.

\subsection{PSF \& Noise Image}
\label{ssec:psf}

For convolution with the point-spread function (PSF) of the {\em HST}
NICMOS optics, and to fit the central point source of the AGN, we
created PSFs using TinyTim (Version 6.3). Compared to other cameras
onboard HST, TinyTim produces fairly good PSFs for NICMOS (and
especially NIC2) because the PSF is less sensitive to aberrations in
the infrared \citep{kri04}; generally, TinyTim PSFs are considered an
adequate alternative when well-matched stellar PSFs are not available
\citep{kim08}.

To minimize PSF mismatch due to spatial distortion, we simulated PSFs
at the location of the objects.  We created a library of 17 PSFs using
a wide range of different stellar templates (F6V to K7V) and power-law
functions ($F_{\lambda} \propto \lambda^{\epsilon}$ with $\epsilon=-3$ to
$\epsilon=0.5$ in increments of 0.5) at the four different chip
positions of the individual exposures of the science targets.  These four
images were then dither-combined using the same procedure as for the
science targets.  To account for breathing, we additionally created
PSF models for the above range with focus values of $\pm$5 $\mu$m
around the nominal focus \citep{rho07}.

We created noise images by dither-combining the necessary extension
files provided in the image block (the output of the CALNICA pipeline;
see NICMOS data handbook Version 7.0) in the same way as the
associated science image.

\subsection{Fitting Procedure \& Uncertainties}
\label{ssec:fit}

For each object, we assumed the following AGN/host galaxy fitting
procedure using GALFIT.\footnote{Note that this procedure is in agreement
with the one used in Paper II and the comparison is therefore
straightforward.}  We first fitted the central AGN component with a
PSF, and thus determined the center of the system, which was
subsequently assumed to be common to all components and fixed.  We
then modeled the spheroid component with a \citet{dev48} profile. We
carefully checked the images and the residuals for evidence of a disk
component and added an exponential disk if required by the images,
residuals, and the $\chi^2$ statistics.  The same approach was used to
determine the need for an additional bar component, but unlike in
Paper II (where seven out of 17 objects required the fitting of a
bar), we did not find evidence for a bar in any of the objects in our
sample.  Neighboring objects were fitted simultaneously.  Note that
the sky was determined independently and subtracted out during
pipeline reduction (see above) which is preferable when using GALFIT,
as the sky background is not only degenerate with the extended wing of
the galaxy, but it might also be used by GALFIT to compensate a
mismatch between intrinsic and fitted galaxy profile
\citep{pen02,kim08}.

To ensure that the best resulting fit indeed corresponds to the true
global minimum of the $\chi^2$ over the parameter space, we performed
a careful inter-comparison between different fits with a variety of
initial parameters and combinations of components.  Finally, each
galaxy was fitted with all PSFs in the library to find the
best-fitting PSF and thus, the best-fitting parameters, and to
estimate uncertainties due to PSF mismatch.  The differences in
derived spheroid and PSF magnitudes using different PSFs in our
library are small ($\le 0.05$ mag) and negligible compared to other
systematic errors.  To understand these systemic errors, we simulated
artificial images spanning the same parameter space as our objects and
tested how reliably GALFIT can retrieve the different components (see
Appendix~\ref{sec:mc}). We use the results to estimate our
uncertainties and adopt a conservative total uncertainty on the
spheroid luminosity of 0.5 mag (i.e., 0.2 dex). The AGN luminosity is
uncertain to within 0.2 mag. The dominating error when constructing
the BH mass - spheroid luminosity relation is the uncertainty of BH
masses from single-epoch measurements, $\sim$0.4 dex.

A PSF+spheroid decomposition gives a satisfactory fit to the host
galaxies of ten out of the 23 objects (see Table~\ref{results}).  The
remaining 13 objects show evidence for a disk component in both the
residual image and in the $\chi^2$ statistics and thus, an additional
exponential disk component was added.  However, in ten of the 13
objects, the addition of a disk component results in a vanishingly
small spheroidal component. In these cases, we fixed the spheroidal
half-light radius to the minimum resolvable size of 2.5 pixels
($\sim$0.1\arcsec), as determined by simulations, and consider the
measured spheroid luminosity an upper limit.  
For one object (1501), even fitting a single spheroid component had the same effect.\footnote{Note, 
however, that fitting this source
is particularly complicated as it is in the process
of merging with a neighboring galaxy.}
Below, we discuss how we
combined these upper limits with informative priors on the
spheroid-to-total luminosity ratio for galaxies to estimate spheroid
luminosities.

Note that compared to the ACS images studied in Paper II,
NICMOS images are less sensitive to a potential disk component,
dominated by young, blue stars.\footnote{At the same time,
NICMOS images have the advantage of being less affected by dust.} We cannot exclude
to have missed a disk component in some of the objects
for which we only fit a spheroidal component. However, this is a conservative
approach, i.e.~reducing any potential offset in the \mbh-\ls~relation.
This is true in general for our procedure: 
We only fit
a two component model consisting of disk+bulge, if there is irrefutable evidence
for a disk component (see e.g.~Fig.~\ref{nicmos}). 
Without such evidence, using only one component
is conservative in the same sense above.

\begin{figure*}[h!]
\includegraphics[scale=0.8]{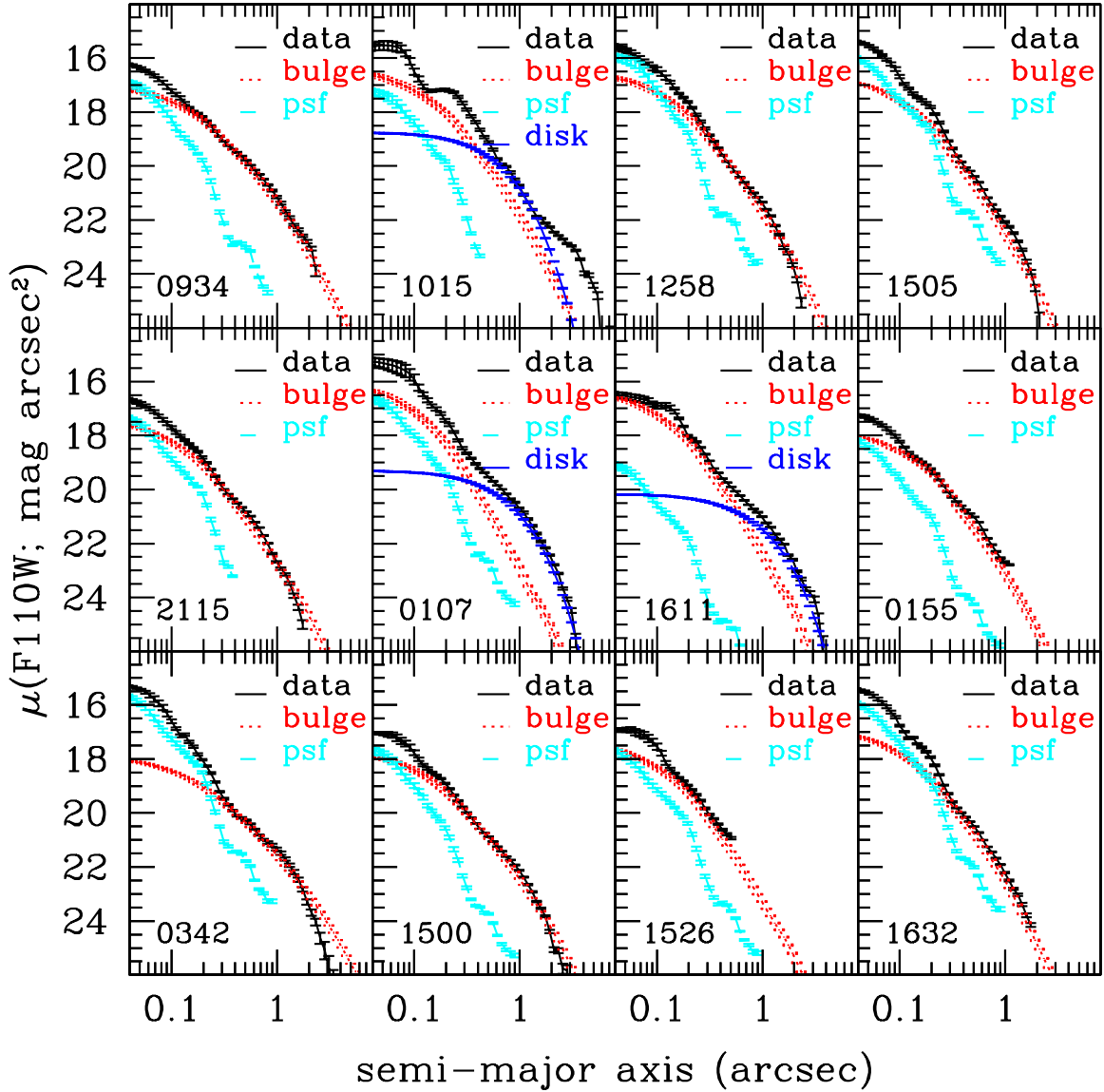}
\caption{Surface-brightness profiles for all
objects with a resolved spheroid component,
measured from the data
as well as from each component that was fitted.
Note that the fits were performed in two dimensions
using GALFIT, so this figure is for illustration purposes only,
showing the relative contribution of each component as a
function of radius. Some profiles show an early truncation
which is an artifact of the elliptical isophote routine used to
make the plots due to nearby objects. 
(For the measurements, these objects were
fitted simultaneously using GALFIT.)
[{\sl See the electronic edition of the Journal
for a color version of this figure.}] }
\label{profiles1}
\end{figure*}

\begin{figure*}[h!]
\includegraphics[scale=0.8]{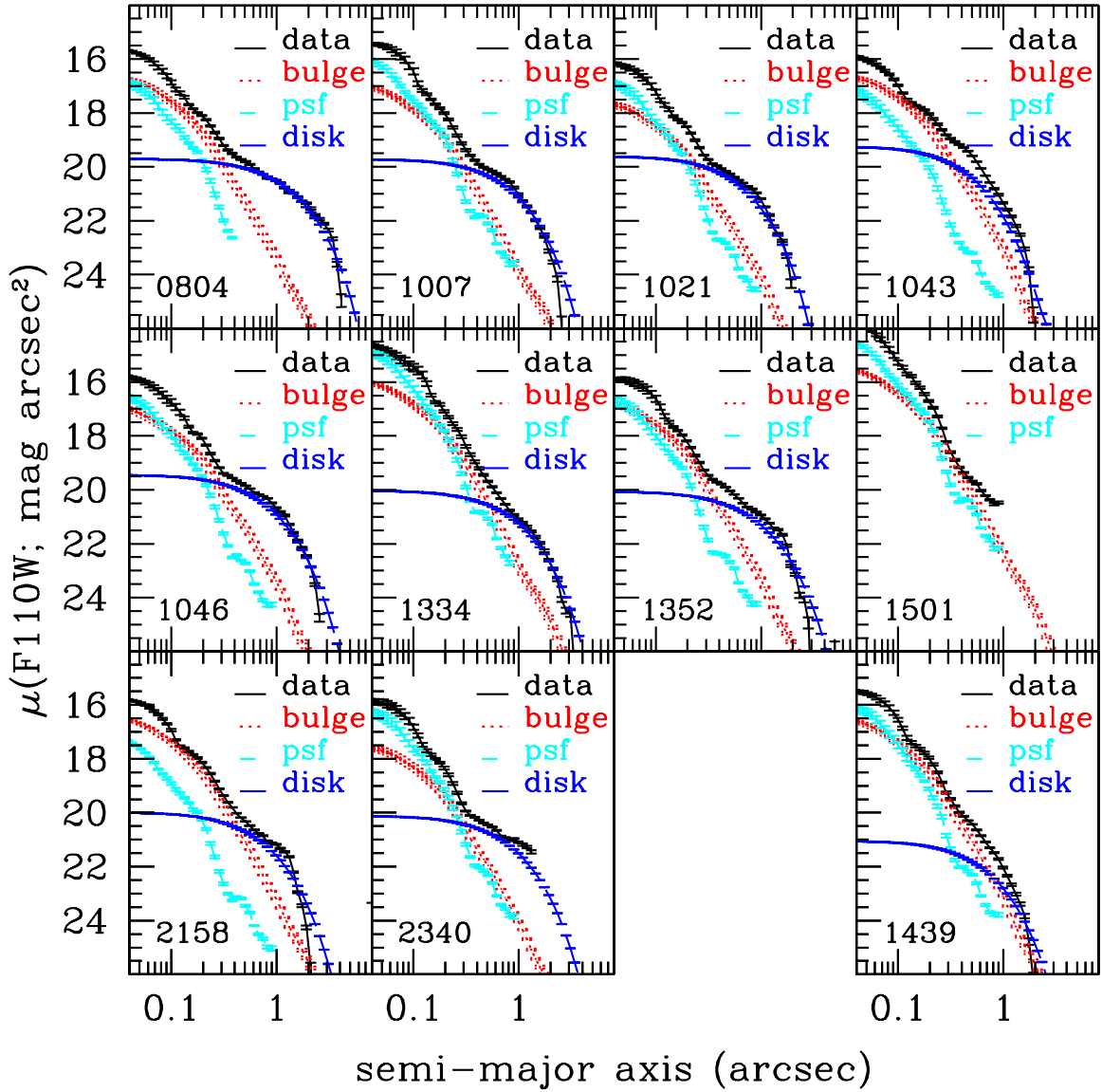}
\caption{The same as in Figure~\ref{profiles1} for 
objects for which
the bulge models correspond to the
minimum size allowed by HST resolution.
[{\sl See the electronic edition of the Journal
for a color version of this figure.}] }
\label{profiles2}
\end{figure*}

In Figs.~\ref{profiles1} and~\ref{profiles2}, we show
surface-brightness profiles derived using the IRAF program
``ellipse'', for the data as well as each component that was fitted
using GALFIT. As the fitting was done in two dimensions, these figures
should be considered as illustrations only, showing the relative
contributions of the different components to the total fit as a
function of radius.  We divide the sample in two groups, based on the
quality of the fit: In Fig.~\ref{profiles1}, we show all objects that
were fitted by a resolved spheroid component; in Fig.~\ref{profiles2},
we show objects with an unresolved spheroid component, i.e.~those for
which GALFIT ran into the size limit of the spheroid of 2.5 pixels.

In Appendix~\ref{sec:sersic}, we discuss the effects
of the choice of different S{\'e}rsic indices other than $n$=4 
(i.e.~a \citet{dev48} profile). To briefly summarize,
all results stated in the paper remain the same within the errors,
when choosing the best-fitting  S{\'e}rsic index instead of $n$=4.

\subsection{Estimating Spheroid Luminosities from Upper Limits Using Informative Priors}
\label{ssec:prior}

As described above, for ten objects the addition of a disk component
resulted in a vanishingly small spheroidal component.
For one object (1501), even a single spheroid component had the same effect.\footnote{As
pointed out before, fitting this source
is particularly complicated as it is in the process
of merging with a neighboring galaxy.}
Thus, for these 11 objects, we fixed
the spheroidal half-light radius to the minimum resolvable size of 2.5
pixels and inferred an upper limit to the bulge luminosity. The same
is true for five objects in Paper II and we thus include them in the
analysis described here.  In brief, we combine the upper limit on
spheroid magnitude with prior knowledge on bulge-to-total luminosity
ratios (B/T) as a function of total host-galaxy magnitude.  In terms
of Bayes' Theorem, we derive the posterior on B/T by combining our
likelihood -- in the form of a step function limited to the measured
upper limit from GALFIT -- with a prior taken from the literature.

The prior is determined from quantitative measurements of the
distribution of spheroid-to-total luminosity ratios
\citep{ben07}, derived for a sample of $\sim$8800 galaxies from SDSS,
using the 2D fitting code GALACTICA \citep{ben02}.  The galaxy
redshifts span a range of 0.02 $<$ z $<$ 0.3 with an average of 0.09.
The absolute R magnitudes of these galaxies are comparable to our
sample (80\% of the \citet{ben02} galaxies are within our range of
$-20.2 \le R_{\rm mag} \le -22.6$).\footnote{Note that we do not
correct for any evolution in luminosity here, as the effect is within
the errors.}

\begin{figure*}[h!]
\includegraphics[scale=0.8]{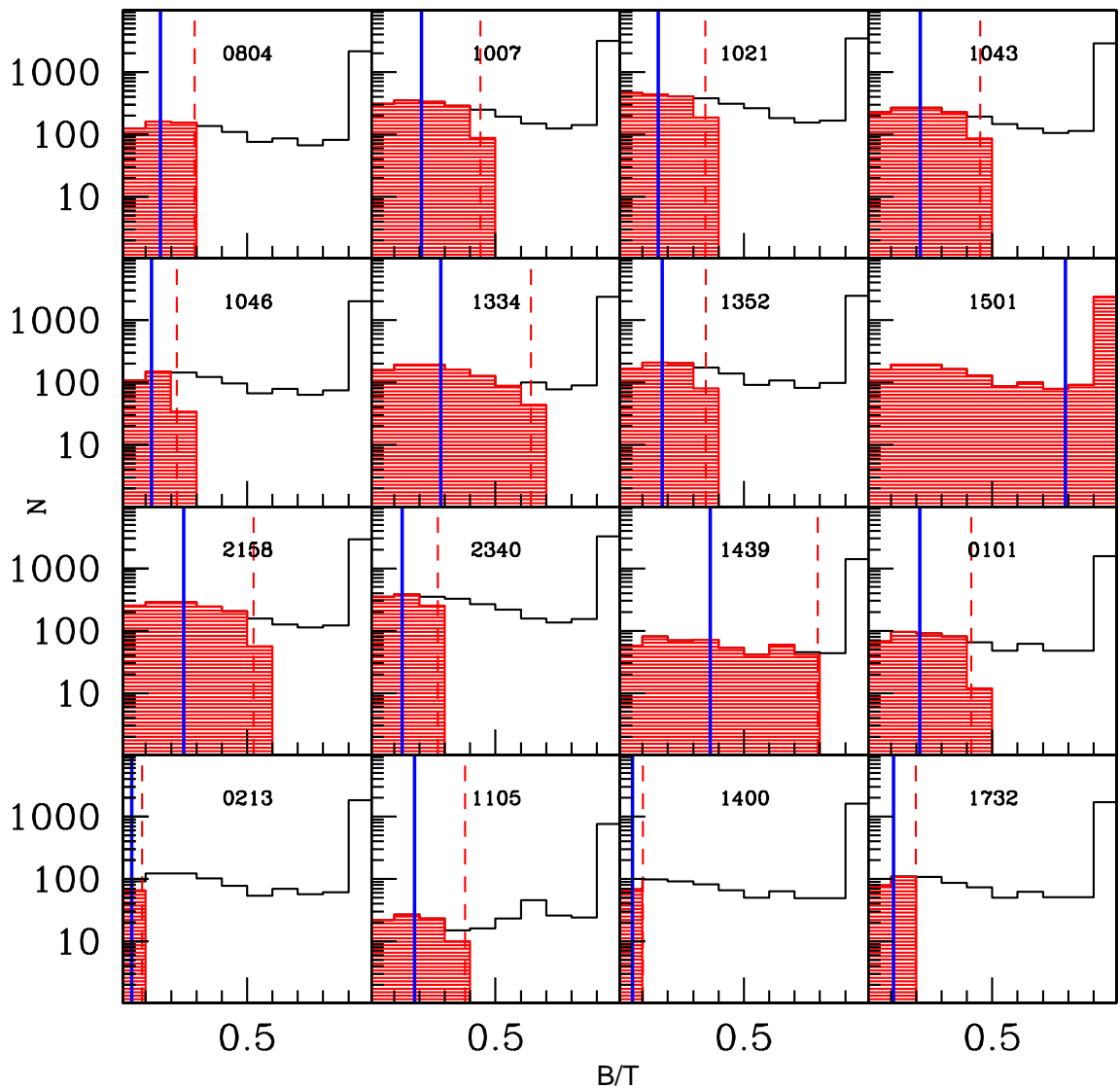}
\caption{Histogram of the bulge-to-total flux ratios (B/T)
from \citet{ben07} for galaxies within 0.5 mag of the
Seyfert total host-galaxy magnitude (black line).
The red dashed line shows the upper limit on B/T
we derived for the Seyfert galaxies from GALFIT.
We use this upper limit to cut the distribution
and to calculate a mean (blue line) and sigma of the remaining
B/T values (red shaded area).
The first 11 objects were imaged with NICMOS
(ten at z=0.36, one at z=0.57)
and the last 5 objects were studied in Paper II,
but we include them here to estimate spheroid luminosities
from upper limits.
[{\sl See the electronic edition of the Journal
for a color version of this figure.}] }
\label{prior}
\end{figure*}

For each object in our sample, we performed the following steps.
First, we selected only objects from \citet{ben07}, for which the
total galaxy magnitude is within $\pm$0.5 mag of the Seyfert total host-galaxy 
magnitude (which is typically accurate to $\lesssim$ 0.1 mag).  
Then, we created a histogram over the B/T values of the Benson et
al. galaxies within this magnitude range (step size of 0.1) and cut
this histogram at the upper limit of B/T we derived for the Seyfert
galaxies.  Finally, we calculated the mean and sigma of the remaining
B/T values and used this value to derive the spheroid luminosity for
the Seyfert galaxies.  In Fig.~\ref{prior}, we show the prior,
likelihood, and posterior distribution functions for all 16
objects. The upper limit and mean posterior B/T ratio is also shown.

\subsection{Dust Correction}
\label{ssec:dust}

For two objects (0107 and 1015), the ACS F775W images studied in Paper
II reveal dust lanes in the host galaxy, preventing an accurate
measurement of the AGN and spheroid luminosity from the ACS images
alone. Thus, NICMOS F110W images were obtained, which we use here for
dust correction. Briefly, the color excess is measured from the two
colors and used to correct for dust extinction assuming an extinction
law. The procedure we adapt is similar to the one described in
\citet{koo03}.

First, we deconvolved the ACS image using the ACS PSF from Paper II
and a Lucy-Richardson algorithm (IRAF program ``lucy'').  Then, the
F775W image was rotated to match the orientation of the NICMOS image
and drizzled to the same pixel scale.  We assume that the centroids
are unaffected by the dust lane -- an assumption supported by the
distribution of the dust seen in the images -- and we thus centered
both images on their peaks.  In the next step, the F775W image was
convolved by the NICMOS PSF (IRAF program ``imconvolve'') to match the
resolution of the F110W image.  Then, a color map was created from the
ratio of these matched images and the intrinsic color was assumed to
correspond to the minimum of the color and to be spatially uniform.
Finally, the color excess was converted into extinction assuming $A_V =
3.1 E_{B-V}$, $A_{\rm F110W}/A_V = 0.628$ (corresponding to rest-frame
F814W), and $A_{\rm F775W}/A_V = 1.049$ (corresponding to rest-frame
F555W).  The extinction-corrected NICMOS image was used for fitting
with GALFIT.

\section{DERIVED QUANTITIES}
\label{sec:quan}

\subsection{Rest-Frame V-Band Luminosities}
\label{ssec:rest}
We applied correction for Galactic extinction, assuming $A_V = 3.1
E_{(B-V)}$ and $A_{F110W} = 0.902 E_{(B-V)}$ \citep{sch98}. The values
for $E_{B-V}$ were taken from \citet{sch98}.  The F110W AB magnitudes
were transformed to rest-frame optical bands by performing synthetic
photometry on an early-type galaxy template spectrum, a procedure
traditionally referred to as k-correction.  The template spectrum
initially has arbitrary units, and these units were adjusted so that
the synthetic observed frame F110W magnitudes match the magnitudes
from our photometry.  We then evaluated the V-band magnitudes at the
rest-frame of the template; luminosities were determined by correcting
for the distance modulus given our adopted cosmology.  The errors on
extinction and rest-frame transformation are a few hundredths of a
magnitude.  We note that the F110W band roughly corresponds to the R
and I bands for our two intermediate-redshift samples; considering the small
scatter in the red colors of bulges (that is, the V-R and V-I colors)
we are able to determine robust estimates of the V-band magnitude.
We estimate an uncertainty of $<$0.05 mag (using the
scatter in 20 single stellar population templates
with ages ranging from 2 Gyr to 8 Gyr).

\subsection{Luminosity Evolution}
\label{ssec:evo}

To allow a direct comparison of the observed relation in the more
distant universe and local samples, we evolved the spheroid luminosity
according to the evolution measured from the evolution of the
fundamental plane by \citet{tre01}:
\begin{equation}
\log L_{V, 0} = \log L_V - (0.62 \pm 0.08 \pm 0.04) \times z
\label{eq:evo}
\end{equation}
This corrects pure passive luminosity evolution, i.e.~the decrease in
spheroid luminosity due to an aging stellar population.  We used the
same correction for our intermediate-z Seyfert sample and for the
local RM AGNs which will be discussed in \S~\ref{ssec:lowzcomp}.
However, equation~\ref{eq:evo} is only valid below $z \lesssim 1$ and an
equivalent measurement is not available at higher redshift. Therefore,
for the high-z sample we adopt a conservative correction based on the
predicted evolution for a maximally old stellar population (see
\S~\ref{ssec:highzcomp}).

\subsection{Black Hole Mass}
\label{ssec:bh}

As in Paper I, II, and III of this series, black hole masses were
estimated using the empirically calibrated photo-ionization method
\citep[e.g.,][for a
detailed discussion see Paper II]{wan99,ves02,woo02,ves06,mcg08}.  
Briefly, the 
method (sometimes called the ``virial'' method) assumes that the
kinematics of the gaseous region in the immediate vicinity of the BH,
the broad line region (BLR), traces the gravitational field of the
BH. The width of the broad emission lines (e.g.~H$\beta$) gives the
velocity scale, while the BLR size is given by the continuum
luminosity through application of an empirical relation found from
reverberation mapping (RM)
\citep[e.g.,][]{wan99, kas00, kas05, ben06}. Combining size and
velocity gives the BH mass, assuming a dimensionless coefficient of
order unity to describe the geometry and kinematics of the BLR
(sometimes known as the ``virial'' coefficient).  This 
coefficient can be obtained by matching the \mbh-$\sigma$ relation
of local active galactic nuclei (AGNs) to that of quiescent galaxies
\citep{onk04, gre06}. Alternatively, the coefficient can be postulated
under specific assumptions on the geometry and kinematics of the
BLR. We note that the exact value of the virial factor does not
affect our results since the relative offset between local and higher
redshift AGNs should be independent of the virial factor.

We use the following formula which includes calibrations
of the BLR size-luminosity relation (after subtraction of
host galaxy light; \citealt{ben06}) and a virial coefficient
taken from \citet{onk04}:
\begin{equation}
\log M_{\rm BH} = 8.58 + 2 \log \left(\frac{\sigma_{\rm H_\beta}}{3000\,{\rm km\,s^{-1}}}\right)
+ 0.518 \log \left(\frac{\lambda L_{5100}}{10^{44}\,{\rm erg\,s^{-1}}}\right)
\end{equation}
with $\sigma_{\rm H\beta}$ the second moment of the broad H$\beta$
emission line and $\lambda L_{5100}$ the observed nuclear luminosity at 5100\AA.

To obtain the 5100\AA~continuum luminosity of the AGN, we
extrapolated the extinction-corrected PSF AB magnitude in F110W to
restframe 5100\AA, assuming the power law $f_{\nu} \propto \nu^{-0.5}$.
The power-law index -0.5$\pm$0.15 is the average derived 
for the majority (24/40) of our sample from SDSS photometry in the restframe
wavelength range 5000\AA-6600\AA~(as covered by SDSS for
z=0.36) after subtraction of host-galaxy light contribution (Szathmary et al., in preparation).
Within the errors, this slope is in agreement with other studies
(see \citealt{van01} and references therein).
The uncertainties on the slope lead to an uncertainty
of $<$10\% in $L_{5100}$ and $<$5\% in $M_{\rm BH}$, i.e.~negligible
compared to the uncertainty of 0.4 dex when deriving BH masses
from single-epoch spectra.

The line width $\sigma_{\rm H\beta}$ was derived from spectra obtained
with the Keck telescope, using the longslit spectrograph LRIS to
measure the stellar velocity dispersion (see Paper I and Woo et al.\
2009, in preparation, for details).  We assume a nominal uncertainty of the BH masses
measured from single-epoch spectra of 0.4 dex.

Note that we do not correct for possible effects of radiation pressure
\citep[e.g.,][see, however, \citealt{net09}]{mar08,mar09}.
First, the role of radiation pressure on
the measurement of BH masses is still discussed
controversially and second, neglecting its effects is a conservative
approach: If radiation pressure does affect the motion of the BLR clouds,
not taking it into account would lead to an underestimation
of the BH mass. Thus, including radiation pressure, the observed
offset would further increase.

All results are summarized in Table~\ref{results}.  In addition to the
23 objects in the sample studied here, we give the results for the
sample in Paper II, which changed slightly ($<$0.15 mag) due to a
small error in the extinction correction in Paper II
(consequently, also the derived BH masses changed slightly).
Also, we determined spheroid luminosities for those objects in Paper II that
only had upper limits by applying the informative prior.  Moreover,
Paper II used the B-band luminosity of the spheroid component (for
comparison with inactive galaxy samples in the local Universe). We
here give the V-band luminosity of the spheroid component.

\section{COMPARISON SAMPLES}
\label{sec:comp}

\subsection{Local Comparison Sample}
\label{ssec:lowzcomp}

Interpreting the \mbh-\ls~relation for the distant Seyfert samples
studied here and any possible evolution with redshift requires a
robust local baseline -- ideally of Seyfert galaxies with comparable
BH masses and spheroid luminosities to avoid selection biases as much
as possible.  The most appropriate local comparison sample for our
study is the reverberation-mapped sample of 35 AGN hosts.  This sample
has the great advantage that the BH mass is derived directly via RM
and does not depend on the BLR size-luminosity relation and its
uncertainties.  A detailed analysis of HST images of the RM sample to
derive AGN and spheroid luminosities was recently completed by
\citet{ben09b}. However, a comparison with our study is not
straightforward due to their very different approach which aims to get
the best estimate of AGN-to-host-galaxy luminosity ratio.  In
particular, while we use the simplest decomposition possible,
i.e.~PSF, spheroid (modeled as simple \citet{dev48} profile) plus
possible exponential disk, \citet{ben09b} allow the S{\'e}rsic index
to vary, sometimes include more than one PSF for the same galaxy (to
compensate for PSF mismatch), and up to three different spheroid
components.

We thus decided to perform an independent analysis of the HST archival
images, for a homogeneous treatment of all the data, using the same
approach as for our distant Seyfert galaxies.  Our robust subsample of
the RM AGNs consists of 19 objects
(10 PG quasars and 9 Seyfert-1 galaxies). 
In Appendix~\ref{sec:rm}, we summarize
the details of the analysis, and show, that our results are in overall
agreement with those of \citet{ben09b}.
However, the spheroid luminosities we derive are often brighter than
in \citet{ben09b}, especially in those cases where we fit a spheroid
component only and not spheroid+disk as in \citet{ben09b},
since we did not find evidence for a disk component.

As for the more distant sample, we correct for passive luminosity
evolution to zero redshift (\S~\ref{ssec:evo}). This is important as
the most massive black holes in the RM sample are systematically found
at higher redshift (up to z=0.29, look-back time $\sim$3.3 Gyrs), 
which changes the best fit
\mbh-\ls~with respect to that presented by \citet{ben09a,ben09b}, who
did not take luminosity evolution into account.  

To compare our local relation to that of \citet{ben09a}, we fitted the data using the
BCES algorithm
\citep{akr96}, which takes into account the effects of errors on 
both coordinates using bivariate correlated errors. Following
\citet{ben09a}, we adopt the bootstrap of the BCES bisector value with
$N$ = 1000 iterations.  We give the different fits in Table~\ref{fits}
in the form of
\begin{equation}
\log \frac{M_{\rm BH}}{10^8 M_{\odot}} = K + \alpha \log \frac{L_{\rm sph, V}}{10^{10} L_{\odot}}
\end{equation}
\label{fit}
However, in line with the Bayesian approach followed in this paper,
instead of using the BCES fitting routine to determine our standard
local baseline, we apply our own fitting routine. Following standard
procedures, gaussian errors on both variables are taken into
account. The intrinsic scatter is a free parameter and is modeled as a
gaussian distribution. Uniform priors are assumed on each free
parameter. The inferred slope and intercept after marginalizing over
the intrinsic scatter are given in Table~\ref{fits}.  As can be
seen, the resulting slope can range from $\alpha$ = 0.67 to $\alpha$ =
0.81, depending on the evolutionary correction and on the fitting
technique. 

For comparison, we transformed the B-band magnitudes
of the local {\it inactive} comparison sample 
from \citet{mar03} (group 1 only) to V-band (assuming an elliptical template
and B-V=0.96 mag; \citealt{fuk95}). Using again our linear fitting routine
including gaussian errors and intrinsic scatter gives
a steeper slope of $\alpha$=1.11$\pm$0.13 (K=0.07$\pm$0.08; scatter=0.38$\pm$0.07).
For a discussion of the difference in slope between AGN
sample and inactive galaxy sample, see \citet{ben09a}.

\subsection{High-z Comparison Sample}
\label{ssec:highzcomp}

To study the evolution of the \mbh-\ls~relation, we selected a
high-redshift comparison sample from \citet{pen06b}, consisting of
a total of 31
gravitationally lensed quasars and 20 non-lensed quasars at redshifts of
0.66 $\leq$ $z$ $\leq$ 4.5.  It includes 15 non-lensed (radio-loud
and radio-quiet) quasars taken from
the literature \citep{kuk01,rid01}. We exclude four objects which were
also excluded from the analysis in \citet{pen06b}, one object, for
which the BH mass is only a lower limit\footnote{Assuming
Eddington-limited accretion} (PSS 2322+1944), as well as two extreme
outliers in \mbh~with high uncertainties (B2045+265 and HE 2149-2745),
leaving us with a sample of 44 QSOs (17 non lensed and 27 lensed objects).
BH masses were estimated from single epoch spectra using the broad
lines \ion{C}{4}, \ion{Mg}{2}, or H$\beta$. The luminosity was derived
from two-dimensional surface brightness fitting (GALFIT).  Note that
the \citet{pen06b} measurements comprise the total host-galaxy
luminosity instead of spheroid luminosity alone, as only
one component was fitted.  However, there is no
evidence for any of the objects to have two components,
indicating that the host galaxies are ellipticals.
(Note that even if there was a disk component present,
fitting only one component is a conservative approach
in the sense that the offset from the local relation 
is the smallest.)

To allow for a homogeneous treatment of the data, we corrected the BH
mass estimation based on the \ion{Mg}{2} line for normalization
differences using the recipe by \citet{mcg08}. Note that for H$\beta$
and \ion{C}{4}, \citet{pen06b} used a comparable normalization factor
and the difference is negligible. For objects for which both
\ion{C}{4} and \ion{Mg}{2} measurements are available, we use the latter line,
as determining \mbh~based on the former may have larger uncertainties:
The \ion{C}{4} line is often found to be blueshifted with
a strong blue excess asymmetry indicating an outflow component
\citep[see e.g.][]{bas05}.

To correct for luminosity evolution, but lacking direct
determinations of passive evolution out to these redshifts, we apply a
conservative evolution correction based on maximally old stellar
populations. Specifically, we assume that the single
burst occurred at $z = 5$. We use a Salpeter initial mass function with
solar metallicity and stellar populations synthesis models from
\citet{bru03} to compute the evolutionary correction.\footnote{We used an Sbc template
for obtaining V-band rest-frame magnitudes (instead
of an elliptical template used for the intermediate- and low-z sample), as it
is closer to the colors predicted by our model.}
Note that assuming a younger stellar population which might
be present \citep[e.g.][]{mar99, can01, eva01, sco03, kau03, san04,
tad05, can06, bar06, jah07} 
-- for example due to triggering of SF from a merger
event that also triggered the AGN activity -- 
we would infer faster
passive evolution and therefore more pronounced evolution in \mbh~at
fixed present-day luminosity. Thus, our approach
is conservative.

\section{RESULTS}
\label{sec:res}

\subsection{Host-Galaxy Morphology And Merger Rates}
\label{ssec:merger}

From the final reduced images (see Fig.~\ref{nicmos}), we derive the
overall host-galaxy morphology. At least five of the 23 objects show a
clear large spiral disk (0804, 1043, 1046, 1352, 2340) and one object
has an extended disk-like structure (1007).  Including the sample from
Paper II, a significant fraction of the host galaxies ($>$15/40) 
have morphologies of Sa or later.

In the NICMOS images, seven objects show evidence for tidal interactions and merging
such as tidal tails and other morphological disturbances (0934, 1021,
2158, 0155, 0342) or nearby companions connected by tidal features
(1501, 1526). Half (3/6) of the objects at z=0.57 show signs
of mergers/interactions (0155, 0342, 1526).  In some cases, it is
difficult to clearly distinguish between the presence of a spiral disk
and tidal tails; we cannot exclude the presence of a tidally disrupted
disk (e.g.~1352, 0342).  Combined with the objects in Paper II, 13 of
40 objects show some sign of tidal disturbance.  When considering only
objects at z=0.36, 10/34 objects are in apparently disturbed systems
(0.29$\pm$0.1).  This agrees with the fraction of disturbed systems
found for a control sample of GOODS galaxies: Selecting all galaxies
within GOODS with comparable stellar luminosity and a redshift range
of $z=0.36 \pm 0.1$ and performing the same visual classification lead
to 12/42 disturbed systems (0.28 $\pm$ 0.08; Paper II). This fraction
is somewhat larger than observed in the local universe
\cite[e.g.,][see, however, \citealt{tal09}]{pat02}. 

\subsection{BH Mass - Spheroid Luminosity Relation}
\label{ssec:mbhl}

\begin{figure*}[h!]
\includegraphics[scale=0.26]{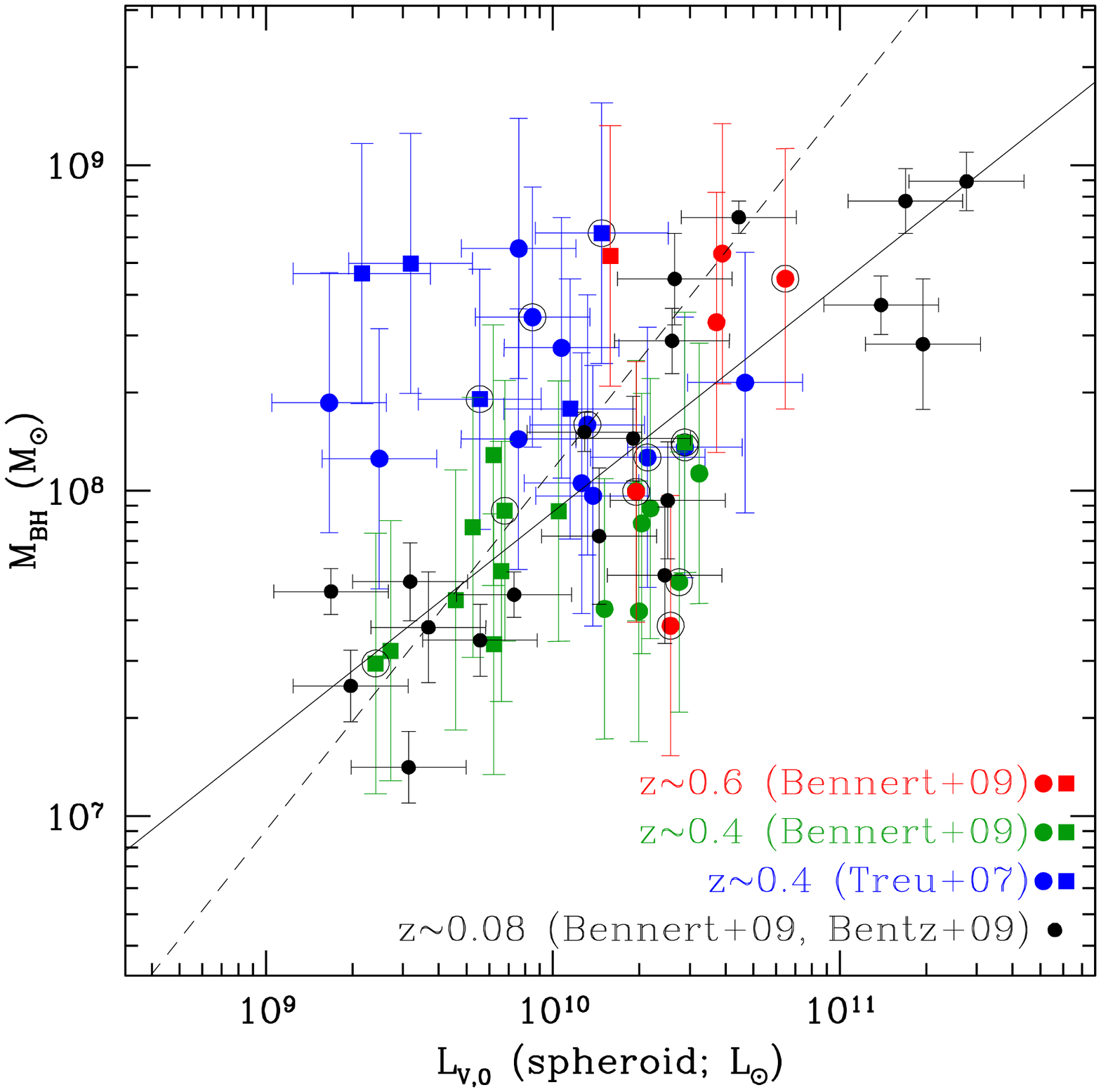}
\includegraphics[scale=0.26]{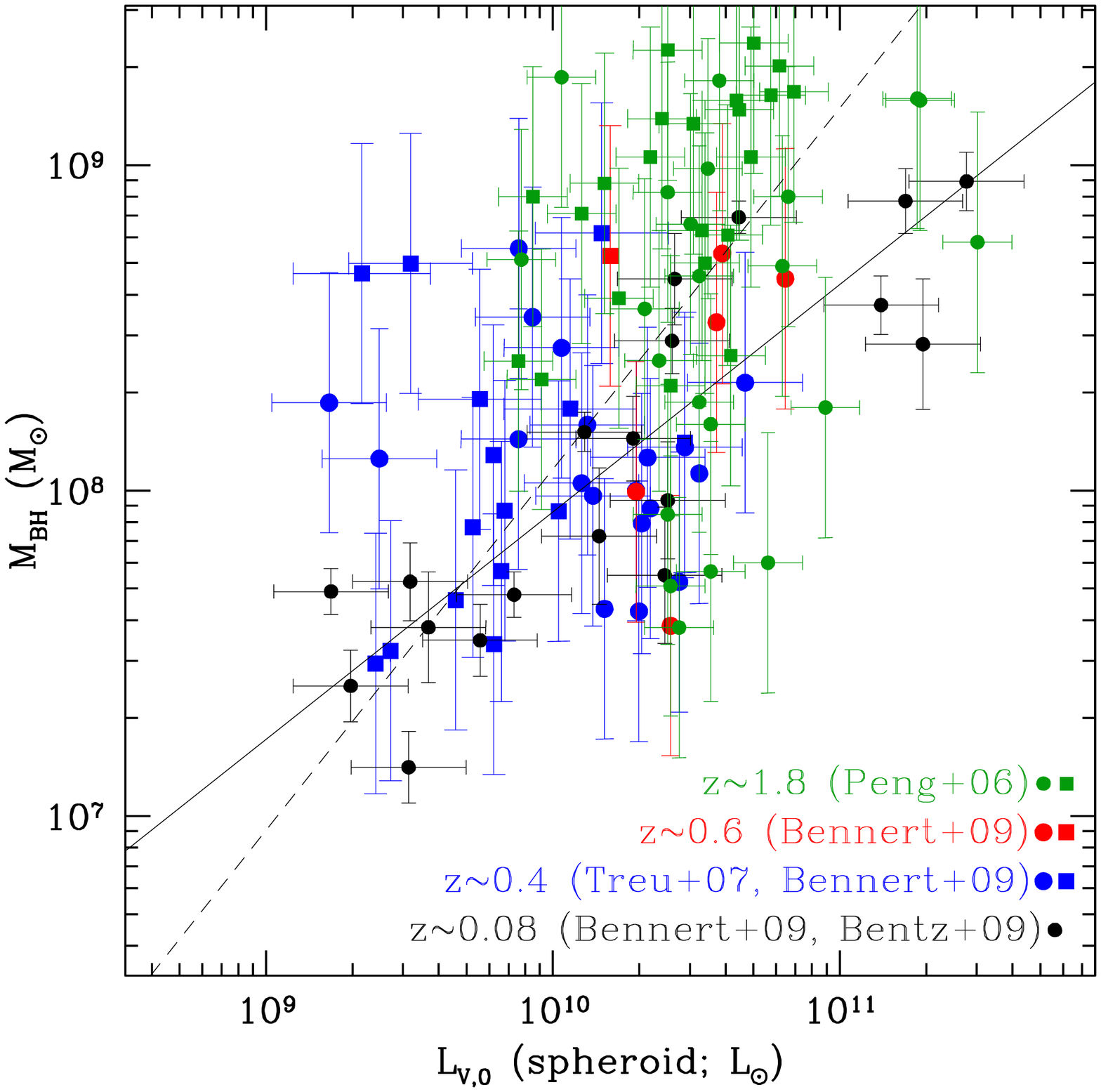}
\includegraphics[scale=0.26]{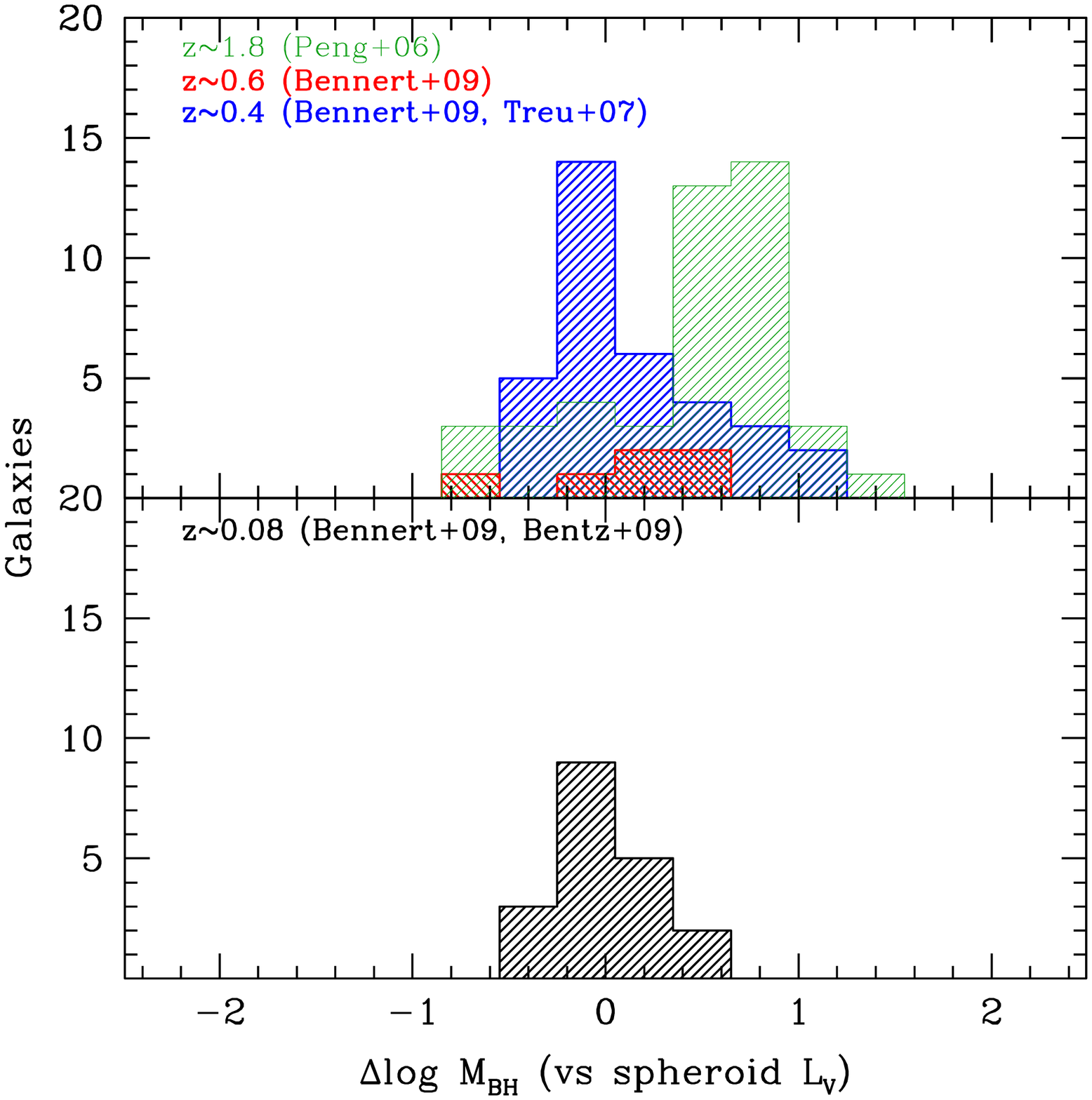}\\
\includegraphics[scale=0.26]{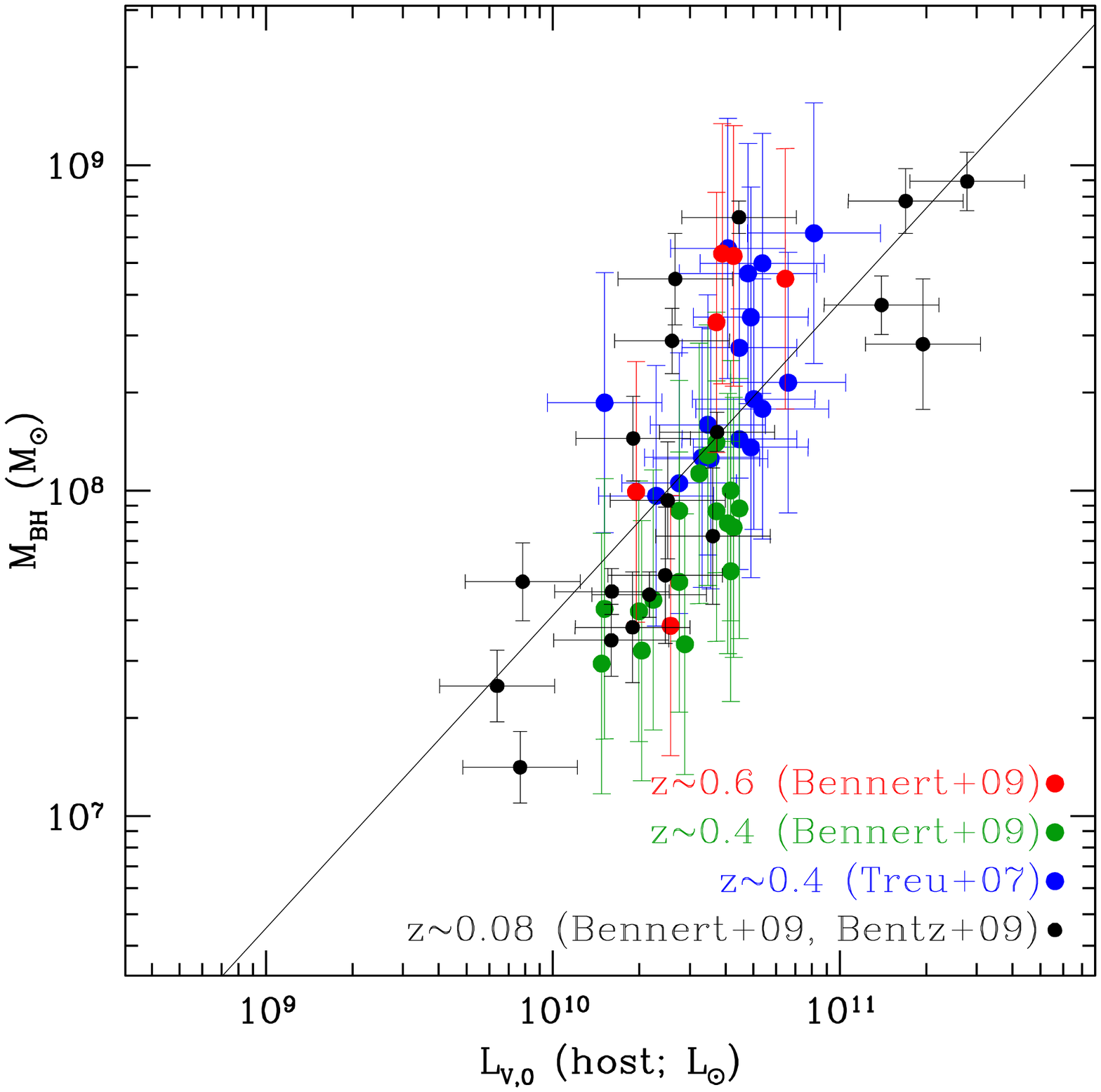}
\includegraphics[scale=0.26]{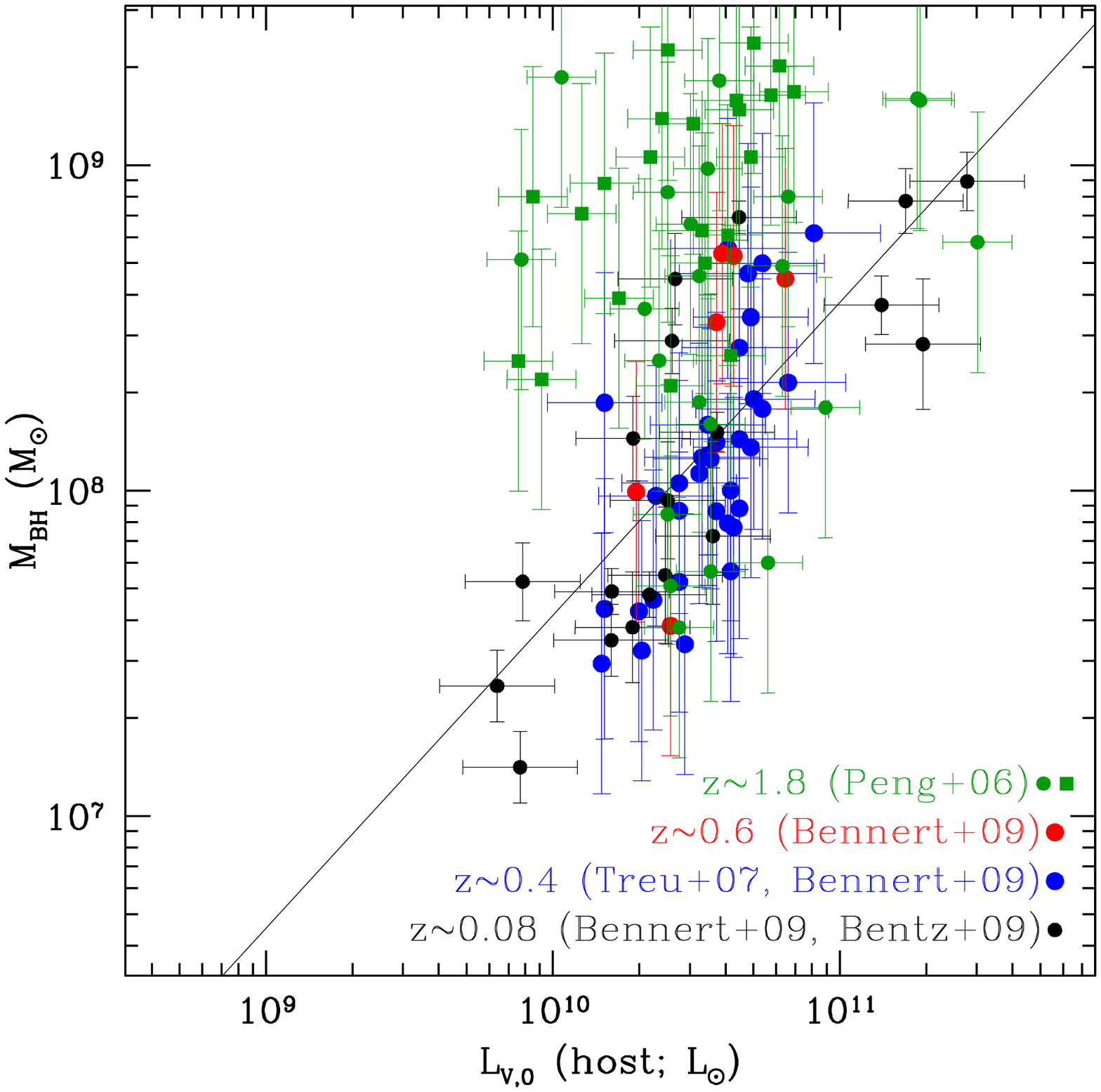}
\includegraphics[scale=0.26]{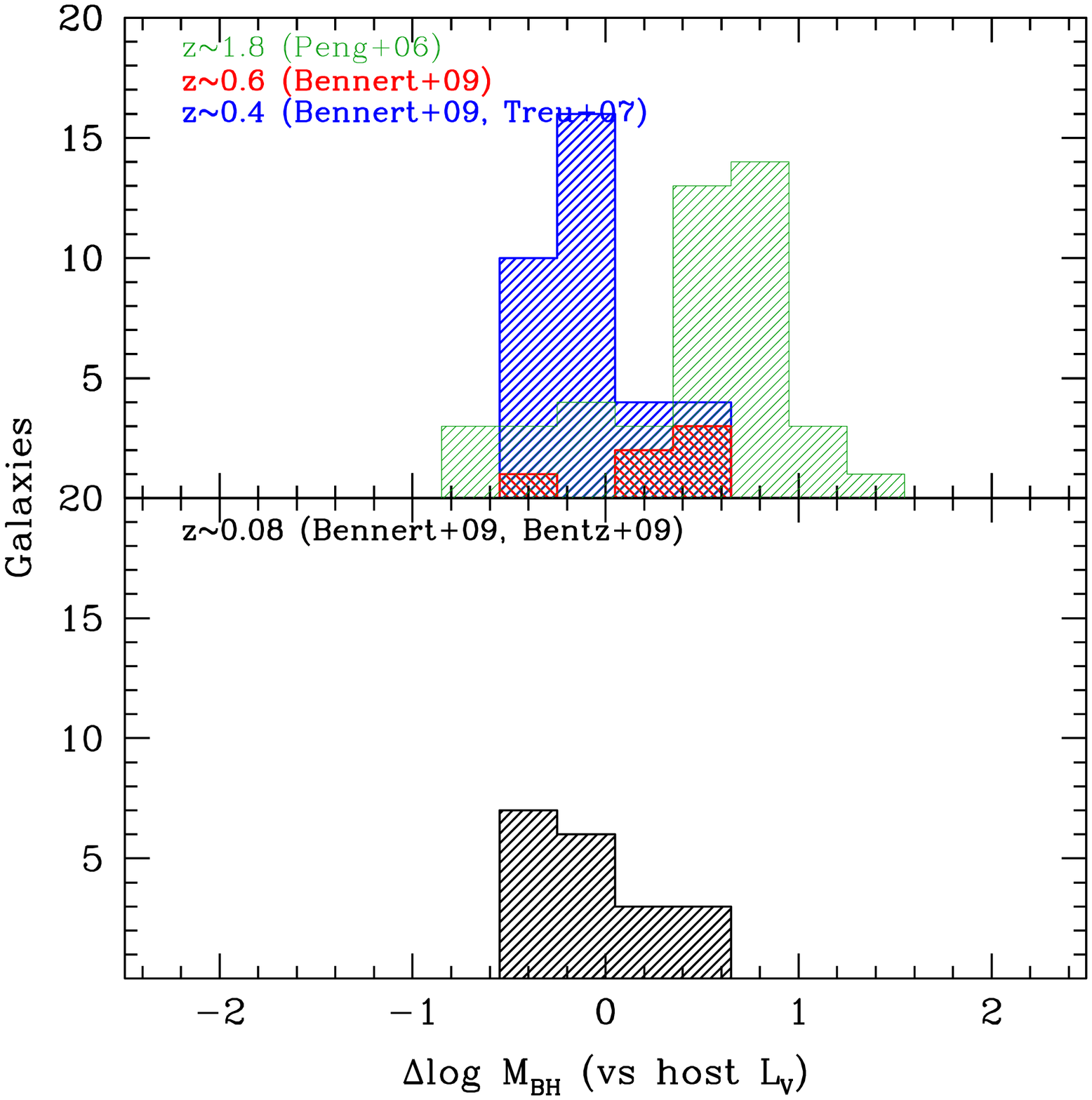}
\caption{
{\bf Upper left panel:} Black hole mass-spheroid V-band luminosity relation.
Colored circles represent measurements for the intermediate-redshift
Seyfert galaxies (red: $z=0.57$, green: $z=0.36$, blue: $z=0.36$ taken
from Paper II; squares indicate objects for which the fitting
procedure ran into the lower limit of the spheroid effective radius
and we used priors to obtain a measure of the spheroid luminosity).
Black circles correspond to the local RM sample ($z_{\rm ave} \simeq 0.08$) studied by
\citet{ben09a, ben09b} and re-analyzed here, including the best fit
(black solid line; see text and Table~\ref{fits} for details). 
For all objects, the spheroid luminosity is evolved to $z=0$ assuming pure
luminosity evolution (see text for details).
Note that no selection effects are included here.
Intermediate-z objects with signatures of interaction or mergers (see Fig.~\ref{nicmos}
and Paper II) are indicated by a large open black circle.
The dashed line shows the fiducial local relation for inactive galaxies \citep{mar03},
transformed to V-band (group 1 only; see text for details).
{\bf Upper middle panel:} The same as in the left panel, this time
all $z=0.36$ objects in blue.
Green circles are the high-z AGN sample (average $z \sim 1.8$)
taken from \citet{pen06b} and treated in a comparable manner.
We assume 0.4 dex as error on \mbh, and 0.12 dex as error on luminosity
(based on the error quoted by \citet{pen06b} of 0.3 mag).
We mark those high-z objects for which the BH mass is based on the \ion{C}{4} line
as green squares.
{\bf Upper right panel:}
Distribution of residuals in log \mbh~with
respect to the fiducial local relation of RM AGNs. 
Top panel: distribution of residuals
for intermediate-redshift Seyfert galaxies (blue: z=0.36; red: z=0.57)
and for the high-z AGN sample from \citet{pen06b} (green). Bottom panel:
local sample. 
{\bf Lower panels:} The same as in the upper panels, 
for the total host-galaxy luminosity. 
 [{\sl See the electronic edition of the Journal for a
color version of this figure.}] }
\label{bhlv}
\end{figure*}

The resulting \mbh-\ls~relation is shown in Fig.~\ref{bhlv} 
(upper left panel).  
Objects with signs of tidal interaction or merger are marked
by open black circles; they are not significant outliers.
In the upper middle panel of Fig.~\ref{bhlv}, the high-z sample is included.
Fig.~\ref{bhlv} (upper right panel) shows the distribution of the
residuals in $\log$\mbh~with respect to the fiducial local relation.
If we treat the intrinsic scatter of the relation as a free parameter
and marginalize over it, the offset we derive with respect to the
local relation (solid black lines in Fig.~\ref{bhlv}, upper left panel)
is $\Delta \log M_{\rm BH}$ = 0.14 $\pm$ 0.07 $\pm$ 0.20
(statistical and systematic error; w.r.t. $L_{\rm sph, V}$ 
including the full sample at both $z=0.36$
and $z=0.57$).  For comparison, in Paper II we found $\Delta \log
M_{\rm BH}$ = 0.51 $\pm$ 0.14 $\pm$ 0.19 (i.e. when considering only
the blue data points in Fig.~\ref{bhlv}, upper left panel). Although the
numbers are consistent within the errors, we note that they cannot be
compared directly due to the radically different selection function in
\mbh, for the two samples (see~\S~\ref{ssec:selection}).
Expressed as offset in spheroid luminosity,
$\Delta \log L_{\rm sph}$ = -0.19 $\pm$ 0.08 $\pm$ 0.21.

\begin{figure*}[h!]
\begin{center}
\includegraphics[scale=0.4]{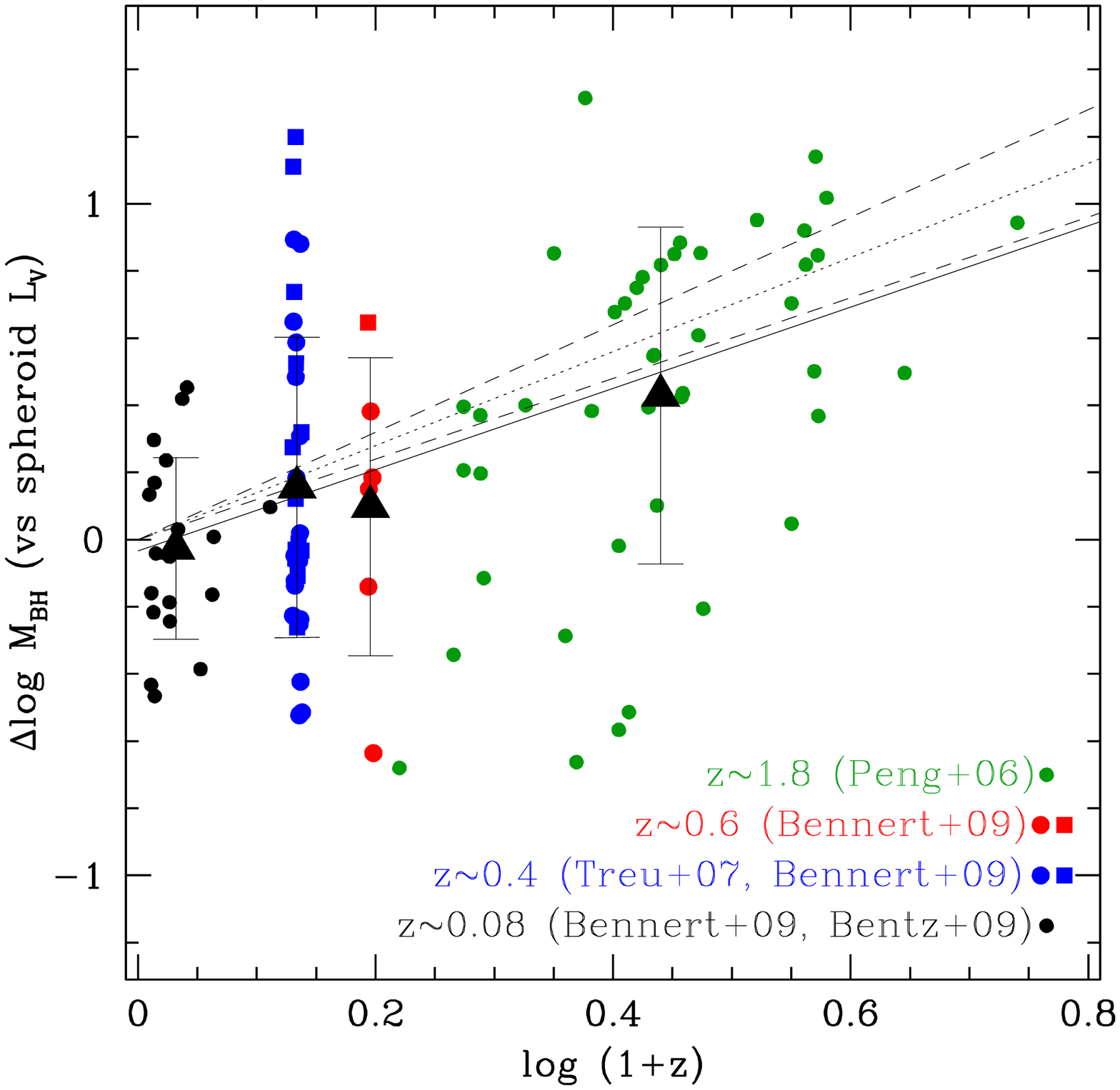}
\includegraphics[scale=0.4]{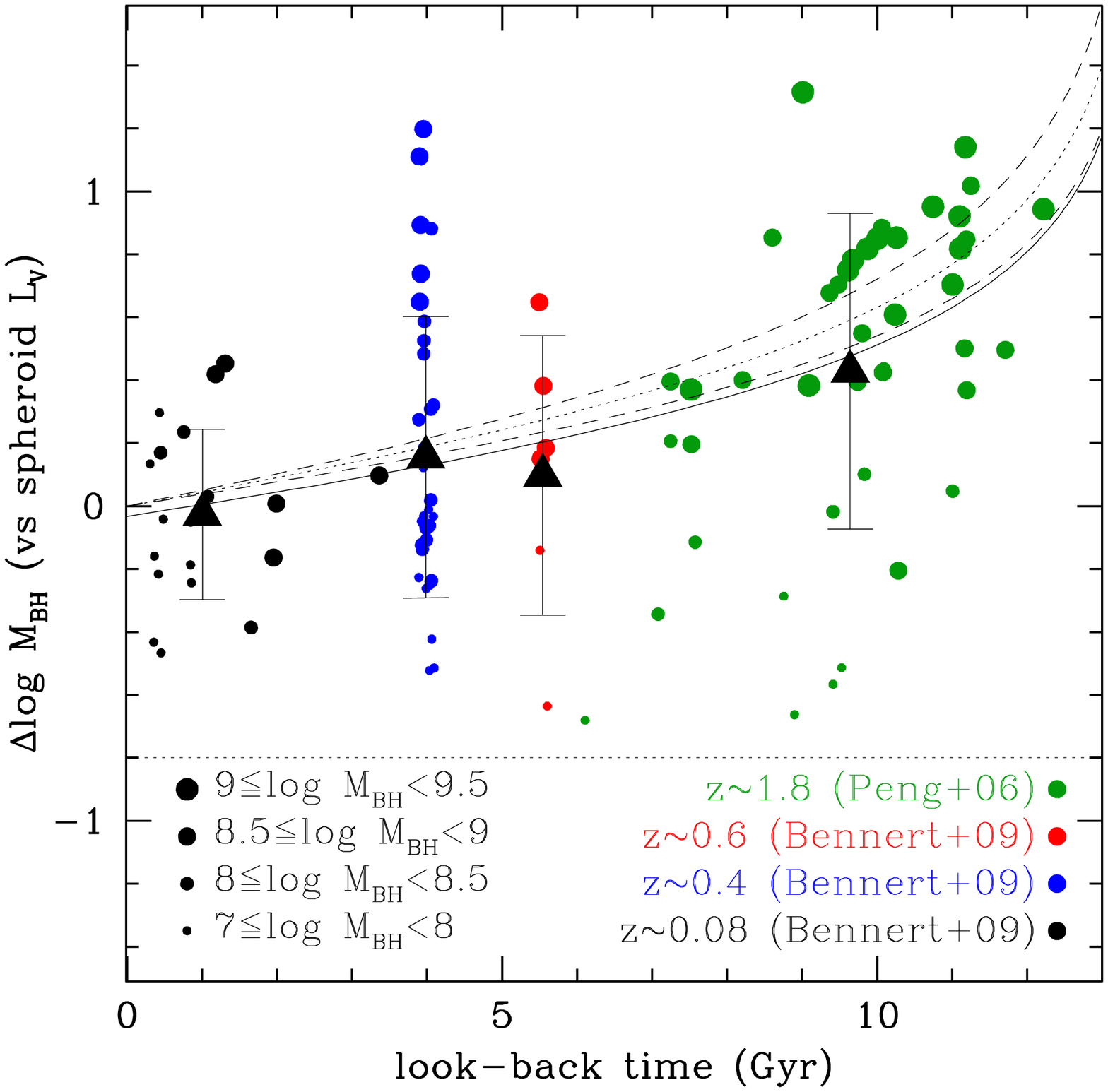}
\end{center}
\caption{{\bf Left panel:} Offset in log \mbh~as a function of log $(1+z)$ with respect to
the fiducial local relation of RM AGNs  (Fig.~\ref{bhlv}, upper middle panel).
The best fit to all data points (solid black line) of the form 
$\Delta \log M_{\rm BH} = \gamma log (1+z)$
including intrinsic scatter in $\log$ \mbh~as a
free parameter but ignoring selection effects is
$\gamma = 1.2 \pm 0.2$.
(Note that the average data points for each sample are plotted only to guide the eye.)
For comparison, we also overplot the selection-bias corrected evolution
(\mbh/\ls~$\propto$ $(1+z)^{1.4 \pm 0.2}$; dotted line) with the 1$\sigma$ range
as dashed lines.
As in Fig.~\ref{bhlv}, squares indicate objects for which the fitting
procedure ran into the lower limit of the spheroid effective radius
and we used priors to obtain a measure of the spheroid luminosity.
{\bf Right panel:} The same as in the left panel as a function of look-back time.
Here, the symbol size corresponds to BH mass. 
}
\label{offset}
\end{figure*}

Studying the evolution with redshift of the offset in BH mass,
with respect to the fiducial local relation,
we fit a linear relation of the form $\Delta \log M_{\rm BH} = \gamma
\log (1+z)$ and include the intrinsic scatter in $\log$ \mbh~as a
free parameter. We assume negligible errors on the redshifts and a
standard error of 0.4 dex for $\log$ \mbh.
We find $\gamma$ = $1.3 \pm 0.9$.  
Note that this fit does not take into account systematic
errors nor selection effects. 
Adding the higher redshift comparison sample from \citet{pen06b},
we extend our redshift baseline, decreasing the error on the slope,
resulting in $\gamma$ = $1.2 \pm 0.2$ (Fig.~\ref{offset}, left panel).
We also plot the evolution as a function of look-back time (Fig.~\ref{offset}, right panel).
However, we stress that the figures and the
fits discussed in this section are for illustration only as they ignore
selection effects. The correct quantitative results taking into
account selection effects are given in~\S~\ref{ssec:selection}.

Note that without correction for passive luminosity evolution,
there is little to no offset for any of the distant objects. 
A similar result has already been found by \citet{pen06b},
who show that host galaxies harboring BHs of the same mass
were as luminous at a redshift out to $z$=4.5 
as they are today, up to $\sim$12 Gyrs later \citep[see also][]{dec09}.

\subsection{Selection Effects}
\label{ssec:selection}
As discussed in Paper II, by selecting targets based on their nuclear
properties and in particular on the presence of a broad line AGN, we
may be biasing our inferred offset \citep[see also][]{lau07}, an effect
analogous to the \citet{mal24} bias.  The
magnitude and sign of the bias depends on the errors, on the selection
function, on the spheroid luminosity function and on the intrinsic
scatter of the correlation with host-galaxy luminosity.  Here, we
exploit the larger sample to correct the offset and infer for the
first time the intrinsic scatter of the relation. The slope is assumed
to be fixed to the local value, given that the dynamic range of the
data is not sufficient for an independent determination of its
evolution.

\begin{figure*}[h!]
\begin{center}
\includegraphics[scale=0.26]{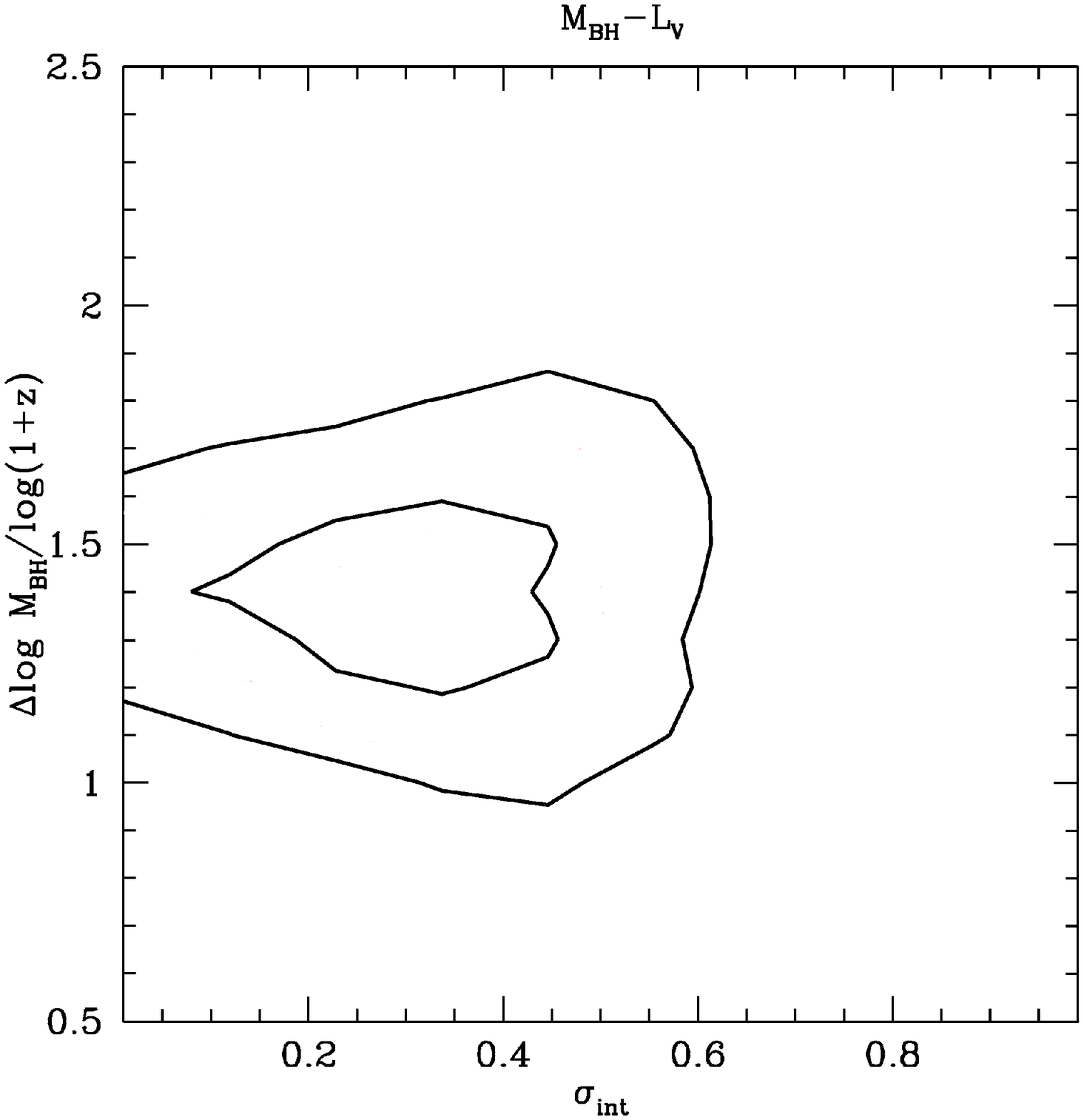}
\includegraphics[scale=0.26]{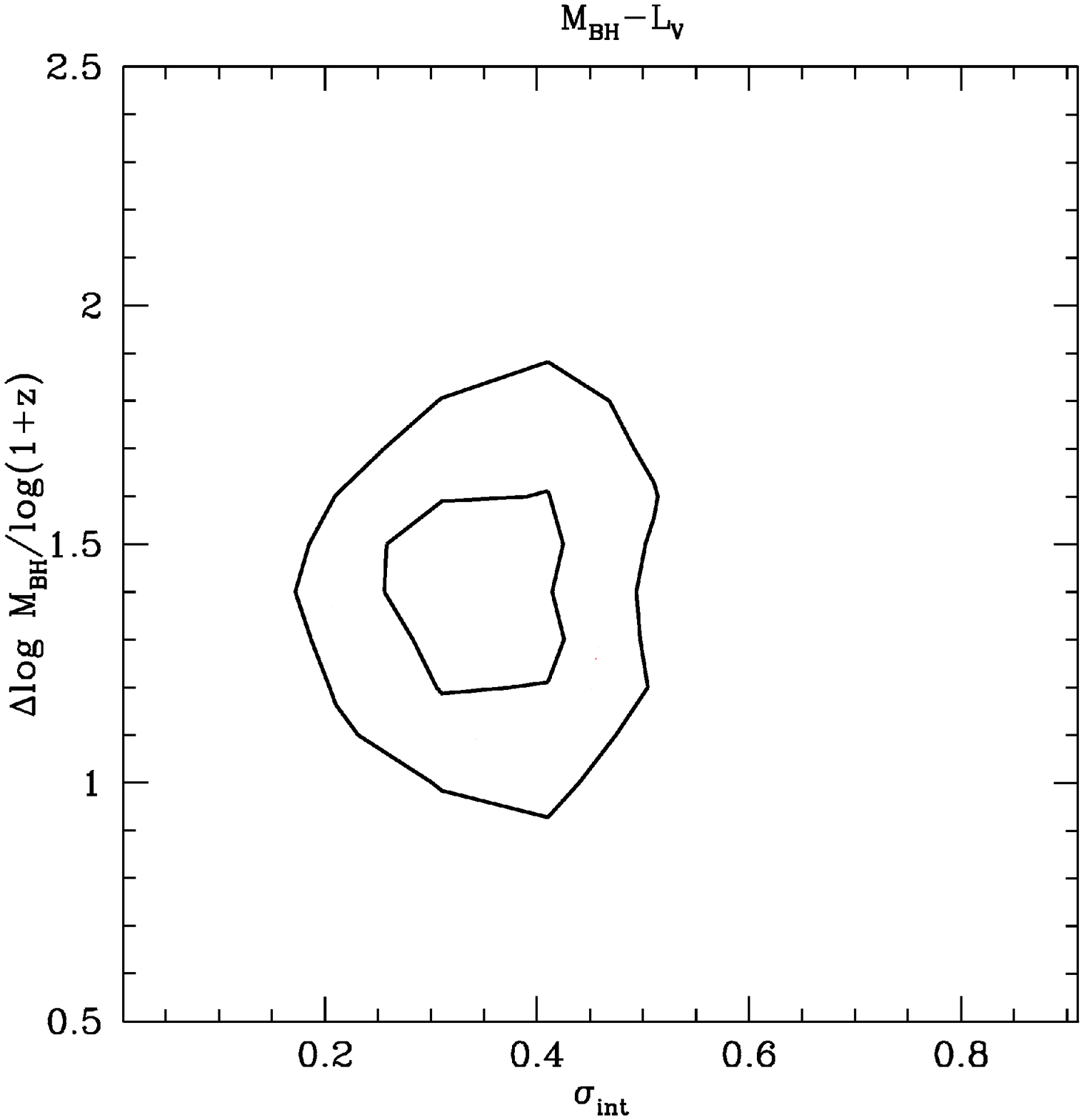}
\includegraphics[scale=0.26]{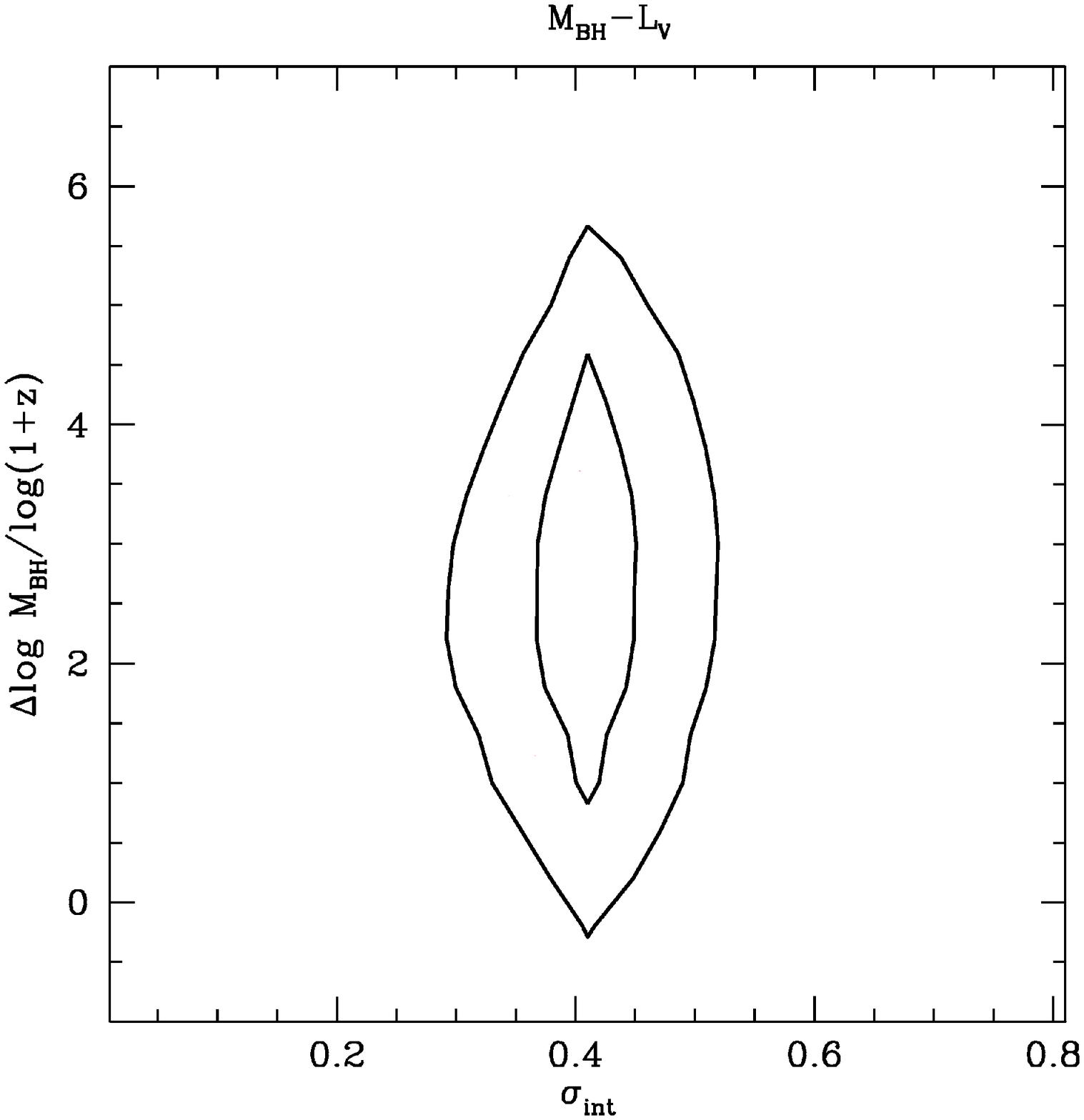}
\end{center}
\caption{Results of Monte Carlo simulations probing the
effect of selection effects on the slope $\beta$ of the relation
$\Delta \log$ \mbh = $\beta \log (1+z)$ at fixed zero redshift
spheroid luminosity corrected for evolution, and intrinsic scatter
$\sigma_{\rm int}$ of the \mbh-\ls~relation which is assumed to be non-evolving. 
Plotted are the 68\% and 95\% joint confidence
contours.
{\bf Left panel:} Including both intermediate-z and high-z sample, without
an assumed prior on $\sigma_{\rm int}$. Both $\beta$ and $\sigma_{\rm int}$ are 
well constrained ($\beta$ =
1.4 $\pm$ 0.2; $\sigma_{\rm int}$ = 0.3 $\pm$ 0.1). 
{\bf Middle panel:} The same as in the left panel,
including the prior by \citet{gue09} (i.e.~$\sigma_{\rm int}$ = 0.38$\pm$0.09),
resulting in the same $\beta$ within the errors.  
{\bf Right panel:} The same as
in the middle panel, but for intermediate-z sample only. While
our sample alone does not cover a large enough range in redshift,
we find $\beta$=2.8$\pm$1.2
using the prior by \citet{gue09} on $\sigma_{\rm int}$.}
\label{selection}
\end{figure*}

Briefly, we use a Monte Carlo approach to simulate the observations
including selection effects and compute the likelihood and posterior
distribution function as a function of the two free parameters: slope
$\beta$ of the relation $\Delta \log$ \mbh = $\beta \log (1+z)$ at
fixed zero redshift spheroid luminosity, and intrinsic scatter
$\sigma_{\rm int}$ of the \mbh-\ls~relation which is assumed to be non
evolving. First, we populate the local \mbh-\ls~correlation according
to the spheroid luminosity function taken from \citet{dri07} (their
Table 1, Sample ``Ellipticals + bulges''). Second, for each value of
the free parameters $\beta$ and $\sigma_{\rm int}$, and for each object in the
distant sample, we generate a simulated observed sample, assuming
gaussian errors on both axes, with amplitude equal to the
observational errors. Third, we model the selection effect by hard
thresholds in $\log$ \mbh, as appropriate for each sample: For the
initial samples of $z=0.36$ Seyferts introduced in Papers I and II,
as well as for the sample at $z=0.57$, we
adopt the interval [7.5,9] (lower and upper value of $\log$ \mbh);
for the $z=0.36$ sample introduced here -- which was selected to have small
black hole masses and was thus restricted
to $\log$ \mbh~$\lesssim$8.2 -- we use [7.5,8.2]; 
for the high-z sample we assume [7.5,10].
It is
important to notice that both the upper and lower limits are relevant
for the analysis, as they bias the results in opposite directions. For
each object in the distant sample, we select simulated objects with
consistent spheroid luminosity within the error, generate a one
dimensional simulated distribution in \mbh, and compute the
likelihood.  To compute the posterior, we adopt a uniform prior on
$\beta$ and two choices of prior for $\sigma_{\rm int}$: i) uniform,
appropriate when the parameter is unknown but believed to be of order
unity; and ii) $\sigma_{\rm int}=0.38\pm0.09$, as determined by \citet{gue09}
for a local sample of inactive galaxies.  The latter is the most
informative prior, although it comes with the price of assuming that
the scatter of the \mbh-\ls~relation is the same for active and
inactive galaxies.  The results are shown in Fig.~\ref{selection}. If
the high-z sample is included, $\beta=1.4\pm0.2$ is well determined
regardless of the assumed prior on $\sigma_{\rm int}$. For a uniform prior on
$\sigma_{\rm int}$, the inferred scatter is 0.3$\pm$0.1 dex
($<$0.6 dex at 95\% CL).  If the
high-z sample is not included, the baseline in redshift is not
sufficient to determine $\beta$ and $\sigma_{\rm int}$ simultaneously for a
uniform prior. Adopting the prior by \citet{gue09} to break the
degeneracy, we find $\beta$=2.8$\pm$1.2.
This is consistent with the trend
observed for the complete sample although the bounds are weaker due to
the smaller baseline in redshift. 

When excluding those 16 objects in our sample for which
we estimated spheroid luminosities from upper limits
using informative priors (\S~\ref{ssec:prior}),
we obtain $\beta$=1.3$\pm$0.3 (including
the high-z sample) and $\beta$=1.8$\pm$1.4 (without
the high-z sample using the same prior as above),
i.e.~consistent within the errors.
If we exclude all objects for which \mbh~was estimated based
on the \ion{C}{4} line (which may be more uncertain; \S~\ref{ssec:highzcomp}),
the evolution is less well constrained, since half of the high-z objects
are excluded.
Using again a prior on $\sigma_{\rm int}=0.38\pm0.09$ as above, it results in 
$\beta$=1.1$\pm$0.3.
The slope also gets shallower when using the local {\it inactive}
galaxy sample from \citet{mar03} (group 1 only, transformed
to V-band magnitude, see \S~\ref{ssec:lowzcomp}):
$\beta$=0.9$\pm$0.2.

\subsection{BH Mass - Host-Galaxy Luminosity Relation}
\label{ssec:mbhtot}

We calculate the total host-galaxy luminosity for both our 
intermediate-redshift Seyfert sample and the local RM AGNs and show the 
\mbh-\lh~relation in Fig.~\ref{bhlv} (lower left panel).
Note that, for
consistency and lack of additional information, we assume the same
k-correction template and passive luminosity evolution for the total
host galaxy as for the spheroid luminosity (see \S~\ref{ssec:evo}).
Conservatively, we also 
assume the same error on the total luminosity as on the spheroid luminosity
of 0.5 mag, although, generally, the error on the total luminosity
is smaller.

Compared to the \mbh-\ls~relation, the \mbh-\lh~relation is apparently non-evolving:
If we again treat the intrinsic scatter of the relation as a free parameter
and marginalize over it, the offset we derive with respect to the
local relation (solid black lines in Fig.~\ref{bhlv}, lower left panel)
is $\Delta \log M_{\rm BH}$ = -0.03 $\pm$ 0.09 $\pm$ 0.04
(w.r.t. $L_{\rm host, V}$; including the full sample at both $z=0.36$
and $z=0.57$).  
Expressed as offset in spheroid luminosity,
$\Delta \log L_{\rm sph}$ = 0.04 $\pm$ 0.09 $\pm$ 0.04.

The
best fit to the local RM AGNs (black solid line in Fig.~\ref{bhlv},
lower left panel) gives a marginally steeper slope than for the
\mbh-\ls~relation ($\alpha=0.96\pm0.18$ vs $\alpha=0.70\pm0.10$;
Table~\ref{fits}).  
Overplotting the high-z comparison sample (Fig.~\ref{bhlv} lower middle panel), 
their luminosity remains the same as in the upper middle panel:
The objects were fitted by \citet{pen06b} by only one component
(without any evidence of a second component) and thus \ls=\lh.
Apparently, the
high-z comparison sample does not follow the same
\mbh-\lh~relation, instead the offset remains.
The distribution of the residuals in $\log$\mbh~of
the distant AGNs with respect to this fiducial local
relation is shown in Fig.~\ref{bhlv} (lower right panel).

\section{DISCUSSION}
\label{sec:disc}

\subsection{The role of mergers}
\label{ssec:mergerd}

Theoretical studies generally invoke
mergers to explain the observed scaling relations between BH mass and
host-galaxy spheroid properties -- a promising way to grow both
spheroid and BH.  In a simple scenario, spheroids grow by (i) the
merging of the progenitor bulges (assuming that both
progenitors have a spheroidal component), (ii) merger-triggered starbursts in
the cold galactic disk, and (iii) by transforming stellar disks into
stellar spheroids
\citep[e.g.,][]{bar92,mih94,cox04}, thus increasing the
spheroid luminosity and stellar velocity dispersion.  The fueling of
the BH, on the other hand, is triggered by the merger event as the gas
loses angular momentum, spirals inward and eventually gets accreted
onto the BH, giving rise to the bright AGN or 'quasar' period in the
evolution of galaxies
\citep[e.g.,][]{kau00,dim05}. Eventually, if BHs are present
in the center of both progenitor galaxies, they may coalesce.  In such
a simple scenario, an evolution in the BH mass - spheroid luminosity
relation is not necessarily expected: Both spheroid and BH grow from
the same gas reservoir, and bulge stars added to the final spheroid
followed the BH mass - spheroid luminosity relation prior to merging,
so the relation will be preserved when the BHs coalesce.  However,
while mergers provide a way to grow both spheroids and BHs, they may
do so on very different timescales.
Moreover, the merger history of
galaxies varies, depending e.g.~on formation time and environment.
Different types of merger, for example with a different relative role
of dissipation \citep[e.g.,][]{hop09a} have different effects on the
growth of spheroid and BH:
For a gas-rich major merger between an elliptical galaxy and
a spiral galaxy - the latter without a (massive) BH -- 
the bulge grows more efficiently than the BH by the disruption of the
stellar disk \citep{cro06}.

In general, our images of the intermediate-z Seyfert
galaxies support the merger scenario (see
Fig.~\ref{nicmos} and \S~\ref{ssec:merger}).  However, 
objects with evidence for merger/interaction do not form any
particular outliers in the BH mass - spheroid luminosity relation
(Fig.~\ref{bhlv}). This may not be too surprising:
For those objects for which we still see two separate galaxies
in the process of merging, we fitted both separately and
the bulge luminosity of the AGN host has not yet increased
from the process of merging.
Other objects with signs of interaction
may be in a later evolutionary stage where the bulge
luminosity has already increased and thus, the object
falls closer to the local relation.
Finally, mergers between similar objects
would only move the system parallel to
the local relation.
In general, the effect of mergers on
the measured bulge luminosity of an object
depends on the type of the merger, the evolutionary stage
of the merger, and the timescales involved
to grow spheroid and BH. 
Such a detailed comparison of merger type and age
with theoretical predictions is beyond the scope of this
paper, given the small sample of merging
objects and the limited information at hand.

Note that the fraction of apparently disturbed systems
we find is not higher than that of a comparison sample of
inactive galaxies at the same redshift (\S\ref{ssec:merger}).
Thus, from our images alone, we cannot infer
a causal link between 
a merger/interaction event and the AGN activity we observe.
Instead, ``normal'' galaxies may have the same merger history,
and ongoing interactions are not necessarily predictive of AGN activity.
The role of mergers for the fueling of AGNs is debated
in the literature 
\citep[e.g.][]{san88,hec84,hut88,dis95,bah97,mcl99,can01,dun03,flo04,can07,urr08,ben08,vei09,tal09}.
While there is little doubt that mergers
are helpful, they are certainly not a sufficient condition,
considering the numerous inactive interacting galaxies.\footnote{However,
this might also be due to the timescales involved, with the
signs of interaction outliving the AGN activity (see also the case of
present-day Type II AGNs; \citealt[e.g.][]{cho09}).}
Also, mergers may be necessary for the high-luminosity QSOs only while 
for Seyfert galaxies, secular evolution
through processes such as bar instabilities may be the dominant
effect in the evolution of these galaxies.
We will come back to this issue in \S~\ref{ssec:mbhtotr}.

\subsection{BH Mass - Spheroid Luminosity Relation}
\label{ssec:mbhlr}

Combining results of low-z, intermediate-z and high-z AGNs, treated in
a self-consistent manner, we can estimate the intrinsic scatter of the
\mbh-\ls~scaling relation and correct evolutionary trends for
selection effects. We discuss scatter and evolution in
the next two subsections.

\subsubsection{Scatter of  \mbh-\ls}
\label{ssec:scatter}

The intrinsic scatter we find (0.3$\pm$0.1 dex; $<$0.6 dex
at 95\% CL) is non-negligible. However, we
assume the intrinsic scatter of the \mbh-\ls~relation
to be non evolving. While it would be desirable to directly study
the evolution of the scatter with redshift, this
requires a larger sample than the one we have at hand.
Actually, we might expect a larger intrinsic scatter at higher redshifts, 
given the different ways and timescales involved when
growing spheroids and BHs through mergers.
Indeed, for the local Universe, the observed tightness in the
relations has been a challenge for theoretical studies.  It has been
explained by self-regulated models of BH growth \citep{hop09b} in which
the energetic feedback of the AGN eventually halts accretion,
preventing the BH from further growth and quenches star formation
\citep[e.g.,][]{cio97,cio01,sil98,mur05,dim05,saz05,hop05,spr05,dim08}. 

Also, a significant fraction of the host galaxies of
both our local RM AGN sample ($\sim$9/19) and
our intermediate-z sample ($>$15/40) are prominent late-type
spirals of type Sa or later which have been found to have a larger intrinsic
scatter than elliptical galaxies \citep[e.g.,][for the
\mbh-$\sigma$ relation: 0.44 dex when including
all galaxies vs. 0.31 for elliptical galaxies only]{gue09}.  
As already discussed in paper II,
the intermediate-z late-type spirals may
eventually fall on the local relation later, through merging, in line with
``downsizing'' \citep[e.g.~][]{cow96,bri00,kod04,bel05,noe07}:
Less massive, blue galaxies merge at later times and
arrive at the local relation by becoming larger, bulge-dominated red
galaxies.  Also, at least some spiral galaxies may not have
classical bulges, but pseudobulges which are characterized by
surface-brightness profiles closer to exponential profiles, ongoing
star formation or starbursts, and nuclear bars or spirals. It is
generally believed that they have evolved secularly through
dissipative processes rather than being formed by mergers
\citep[see e.g. review in][]{kor04}.
BHs have been found to reside in galaxies without classical bulges
which may not follow the same scaling relations
\citep[e.g.,][]{gre08}.

\subsubsection{Evolution of \mbh-\ls~with redshift}
\label{ssec:evor}

To generalize our results and to facilitate comparison with
theoretical and observational works, it is useful to estimate the
evolution of the \mbh - spheroid {\it stellar mass} relation.  We can
convert the observed evolution of \mbh~- spheroid luminosity into that
between \mbh~and spheroid mass, if we assume that -- after correction
for luminosity evolution -- the mass-to-light ratio does not change
from sample to sample\footnote{Unfortunately, spatially resolved color
information for a more sophisticated estimation of the stellar mass of
the bulge is currently not available.}.  Under this assumption, an
offset of $\Delta$\mbh~at fixed \ls~equals that at fixed $M_{\rm
star}$ and thus, \mbh/\ms~$\propto$$(1+z)^{1.4 \pm 0.2}$.

We are now in a position to make a broad range of comparisons.
In the literature, the BH mass evolution is discussed quite controversially.
\citet{shi03} study the \mbh-$\sigma$ relation 
out to $z=3.3$, estimating \mbh~from H$\beta$ and $\sigma$ from [\ion{O}{3}]
and find that the QSOs and their host galaxies follow the local relation.
(Note, however, that using [\ion{O}{3}] as a surrogate for $\sigma$ can
be problematic as [\ion{O}{3}] is known to often have an outflow component;
for a discussion see e.g. \citet{gre05,kom07}.)
A similar conclusion has been reached by \citet{shen08}
who study over 900 broad-line AGNs out to $z \simeq 0.4$ from SDSS.
\cite{ade05} use the correlation length of 79 quasar hosts
at $z \sim 2-3$ to estimate the virial mass of the halo and the \ion{C}{4} line
width and UV flux at 1350\AA~to estimate \mbh.
When comparing the resulting \mbh~-$M_{\rm halo}$ relation
to the local one \citep{fer02}, they find no evidence for
evolution. In particular, they can
rule out evolution of the form \ms/\ms~$\propto$$(1+z)^{2.5}$
with $z=2.5$ at 90\% CL, given their error bars.

Other observational studies 
find the same trend in evolution as we do, i.e.~that BHs are too massive
for a given bulge mass or velocity
dispersion at higher redshifts \citep{wal04,shi06,mcl06,pen06b,sal07,wei07,rie08,rie09,gu09}.
\citet{mcl06}, for example, study radio-loud AGN ($0 < z < 2$)
and find \mbh/\ms$\propto$$(1+z)^{2.07 \pm 0.76}$.
\citet{pen06b}, whose data, treated
in a consistent manner to match our data set, are
included in this study, rule out pure luminosity evolution and
find that the ratio between \mbh~and $M_{\rm sph}$ was
$\sim$four times larger at $z \sim 2-3$ than today.
For 89 broad-line AGNs between $1<z<2.2$ in the zCOSMOS survey,
\citet{mer09} find \mbh/\ms$\propto$$(1+z)^{0.74 \pm 0.12}$.
However, this fit refers to the total host galaxy instead of the spheroid
component alone. At least some galaxies will have a non-negligible
disk fraction, which, when taken into account, would result
in a larger offset \citep[see also][]{jah09}.
For a sample of $\sim$ 100 quasars
selected to reside in elliptical hosts,
\citet{dec09} estimate that 
\mbh/\ms~was $\sim$8 times larger at $z \sim 3$ than today,
i.e.~\mbh/\ms$\propto$$(1+z)^{1.5}$.
The evolution we find
is also consistent with our previous results,
within their much larger errors:
\mbh/\ms$\propto$$(1+z)^{1.5 \pm 1}$ from the \mbh-\ls~relation in Paper II, 
whose data are included here, and $\Delta \log$ \mbh$\propto$$(1+z)^{3.1 \pm 1.5}$
from the \mbh-$\sigma$ relation in Paper III.

From a theoretical perspective, the discussion on the evolution 
of the BH mass scaling relations is not any less controversial.
\citet{sha09}, for example, use the local velocity dispersion function (VDF) 
of spheroids, together with their inferred age distributions, 
to predict the VDF at higher redshifts.
Using  the \mbh-$\sigma$ relation with a normalization allowed to evolve with redshift
($\propto (1+z)^{\delta}$), they infer the BH mass density and compare it to the accumulated 
BH mass density derived from the time integral of the AGN LF.
They find a mild redshift evolution ($\delta < 0.35$),
excluding $\delta > 1.3$ at more than 99\% CL
(with the possibility of a stronger evolution for the more massive BHs).
 Another study using fully cosmological
hydrodynamic simulations of $\Lambda$CDM following the growth of
galaxies and supermassive BHs, as well as their associated feedback
processes, finds only limited evolution in \mbh~with a steepening at
z=2-4 \citep{dim08}.
\citet{mer04} expect a weak evolution of \mbh/\ms~$\propto$
(1+z)$^{0.4-0.6}$, when fitting the total stellar mass and star formation rate
density as a function of redshift and comparing that to the hard X-ray
selected quasar luminosity function, assuming that BHs only grow
through accretion. Such a slope is in agreement with work by \citet{hop09a}
who combine prior observational constraints in halo occupation models
with libraries of high-resolution hydrodynamic simulations of galaxy
mergers. 
Using semi-analytic models, \citet{cro06} predicts
an evolution of \mbh/\ms~$\propto$ (1+z)$^{0.4-1.2}$.
A more rapid evolution is predicted by
\citet{wyi03} who assume a self-regulated BH growth model 
and find \mbh/\ms~$\propto$ (1+z)$^{1.5}$,
similar to our observational result.

However, the great advantage of the study presented here are 
the high-quality images at hand, allowing for a detailed
bulge-to-disk decomposition of the host galaxy of
the low- and intermediate-z Seyfert-1 galaxies.
Combining data from a large sample of active galaxies, covering
a redshift range from the local Universe out to z=4.5,
all treated in a consistent manner, results in smaller error bars
on the predicted evolution  than previous studies.
Moreover, it allows, for the first time, to correct evolutionary trends
for selection effects. 
The evolution we find (\mbh/\ms$\propto$$(1+z)^{1.4 \pm 0.2}$)
is indicative of BH growth preceding host-spheroid assembly.

Still, we did not take into account
that the evolution may depend on BH mass (see also Paper III). 
Indeed, there are theoretical
predictions that objects with higher BH (or bulge) masses evolve
faster \citep[e.g.,][]{hop09a}. For example, \citet{dim08} find that when
restricting their fits to objects with $M_{\star}\geq 5\times 10^{10}\
M_{\odot}$, the relation has a slope of $\sim 1.9$ at z=3-4 and $\sim
1.5$ at z=2.  Unfortunately, our sample is too small 
to allow us to address this possibility. There may indeed be
some evidence that the offset in BH mass is larger for objects with
more massive BHs (Fig.~\ref{bhlv}, upper left panel and Fig.~\ref{offset}, right panel).

\subsection{BH Mass - Galaxy Luminosity Relation}
\label{ssec:mbhtotr}

A different scenario seems to emerge when considering the relation
between \mbh~and {\it total host-galaxy} luminosity (Fig.~\ref{bhlv},
lower left panel).  This relation is almost non-evolving within
the last six billion years.\footnote{Note
that there is insufficient information
to constrain the intrinsic scatter.}
Recently, \citet{jah09} found qualitatively similar
results for a small sample of ten AGNs at redshifts between
1 $< z <$ 2: They derive host-galaxy masses from
colors based on ACS and NICMOS imaging, finding that they
lie on the \mbh-$M_{\star, bulge}$ relation in the local Universe
\citep{har04}.

Such a non-evolving \mbh-\lh~relation can be interpreted twofold.

(a) The amount by which
some of the more distant objects have to grow their spheroid is
already contained within the galaxy itself, and the growth
can be achieved by the redistribution of stars, 
i.e.~transforming disk stars into bulge stars.
Such a redistribution can be the result of mergers or
secular evolution, e.g.~bar instabilities
\citep[e.g.,][]{com81,vanb98,avi05,deb06} and torque-driven accretion
\citep[see e.g. review in][]{kor04} which may coincidentally be
also the triggering mechanism for the BH activity we observe
\citep[e.g.,][]{shl93,ath03,dum07,haa09}.
While not every object in the
intermediate-z sample will experience
a major merger in the last 4-6 billion years,
secular evolution is a promising alternative way to
grow the spheroidal components in these objects.
But even if they do experience a major merger (as indeed
evidenced for at least some objects in our sample),
the role of the merger depends on the merger type
as discussed above (e.g.~a merger between similar objects
will simply move the system along the local relation).
There may again be a dependency on BH mass:
For the low-mass objects, the offset becomes
almost negative, indicating that in the low-mass range, 
either the BH is, at the same time, still growing by a non-negligible amount
(consistent with the higher Eddington ratio in the low-mass regime\footnote{Of course,
we may be biased against low-mass objects with low Eddington ratios.})
or that not all of the stellar mass will end up in the spheroid component.
Indeed, for local RM AGNs, at least 6/19 objects reside in late-type host galaxies
(preferentially those with lower BH masses).

(b) The relation between BH mass and host-galaxy luminosity (or mass) may be
the more fundamental one. Indeed, this is predicted by
\citet{pen07}: In his thought experiment, 
he shows that a tight linear relation between \mbh~and host-galaxy mass can evolve --
if the galaxy mass function declines with increasing mass --
due to ``a central-limit-like tendency for galaxy mergers, which is
much stronger for major mergers than for minor mergers, and a
convergence toward a linear relation that is due mainly to minor
mergers''.
Also, it is possible that BHs in late-type galaxies or
galaxies without classical bulges, while not following
the same \mbh~scaling relations as spheroids (see discussion in \S~\ref{ssec:scatter}), they
instead obey a more fundamental relation between BH mass and host-galaxy mass.

However, the relation between host-galaxy luminosity and
\mbh~seems to exist only up to z$\lesssim$1: The offset for the high-z
comparison sample does not decrease as the luminosity given by
\citet{pen06b} is already the total host-galaxy luminosity
\citep[the same is true also for the results by][]{mer09,dec09}.
Along the line of argument of (a) above, the
growth of the spheroid above a redshift of $z \gtrsim 1$
cannot simply be achieved through secular evolution
(with quasars being predominantly hosted by ellipticals), but
instead, major mergers are needed. A major merger is more
likely to happen for the high-z sample given the longer
time span.
Or, following (b), a relation between BH mass and host galaxy is already at place at
$z \lesssim 1$, but still evolving at earlier times.
However, we cannot exclude that part of the difference
is due to the difference in BH mass between the samples,
with the high-z objects generally having larger BH masses.

In the end, the discussion boils down to
the following question:
What is the dominant mechanism that grows
spheroids, and does it depend on spheroid mass and/or redshift?
This is debated controversially in the theoretical literature.
For example, based on their semi-analytic models, \citet{par09} 
find that the majority of ellipticals and spirals 
never experience a major merger but rather, that they acquire
their spheroid stellar mass through minor 
mergers or disc instabilities.
\citet{hop09c}, on the other hand, combine empirically
constrained halo occupation distributions with 
high-resolution merger simulations,
and find that major mergers dominate the
formation of $\sim$$L_{\star}$ bulges and systems
with higher B/T, but that lower-mass or lower B/T
systems are preferentially formed by minor mergers.
They predict that the major merger rate
increases with redshift.
Qualitatively, we can reconcile such a scenario with our results:
Higher-mass objects and those
at higher redshifts (i.e.~the majority of the high-z sample)
form their spheroids preferentially through major
mergers and are thus still evolving
toward a \mbh-\lh~relation,
while lower-mass and lower-z objects (i.e.~our
intermediate-z sample) grow their spheroids
through minor mergers or disk instabilities that redistribute
the stars and thus, they fall on the \mbh-\lh~relation.

\section{SUMMARY}
\label{sec:sum}

We study the evolution and intrinsic scatter of the BH mass - spheroid
luminosity relation, taking into account selection effects, by
combining three different samples of AGNs.  Our intermediate-redshift
sample comprises 40 Seyfert-1 galaxies at two different redshift bins (34
objects at z=0.36, and 6 objects at z=0.57; look-back time 4-6 Gyr)
for which we measure the BH mass from single-epoch Keck spectra.  
The sample spans more than one order of magnitude
in BH mass (log \mbh/$M_{\odot}$=7.5-8.8).  2D surface-brightness
photometry using GALFIT is carried out on high-resolution HST images
to decompose the image into AGN and host-galaxy components.  The low-z
comparison sample consists of 19 local AGNs ($0.02 \leq z \leq 0.29$; $z_{\rm ave}=0.08$) 
with reverberation BH masses \citep{ben09b}. 
We re-analyzed the archival HST images in a way comparable to our
intermediate-z Seyfert galaxies to eliminate possible systematic
offsets.  Finally, we combine our results with high-z data 
(44 quasars from 0.66 $\leq$ $z$ $\leq$ 4.5; $z_{\rm ave}=1.8$) compiled
from the literature, mainly consisting of gravitationally-lensed AGNs
\citep{pen06b} that were treated in a self-consistent manner.
For all objects, the spheroid luminosity is corrected for passive luminosity
evolution. Our main results can be summarized as follows.

\begin{itemize}
\item{
We determine the evolution in \mbh~with 
an unprecedented accuracy, taking
into account selection effects. Our result,
\mbh/\ls$\propto$$(1+z)^{1.4 \pm 0.2}$,
indicates that BH growth precedes host-spheroid assembly.
The intrinsic scatter, assumed to be non-evolving, 
is non-negligible (0.3$\pm$0.1 dex; $<$0.6 dex at 95\% CL).
It may reflect the different ways and timescales involved
when growing spheroids or may partially be due to a 
high fraction of spirals and/or potential
pseudobulges in our sample.}

\item{
The local and intermediate-z sample follow an apparently non-evolving
\mbh-{\it host-galaxy} luminosity relation.
Either the spheroid grows by a redistribution
of stars, or  the relation between BH mass and host galaxy is 
more fundamental. Above $z \simeq 1$, the relation seems to be still 
forming, e.g.~through major mergers.}
\end{itemize}

We are currently studying the evolution of the BH mass - spheroid velocity
dispersion relation (Woo et al. 2009, in preparation), which
should allow us to tighten the error bars on evolution given
that velocity dispersion can be measured more precisely than host
luminosity. Studying this independent relation will also enable us to
distinguish between different evolutionary scenarios, probe the 
``fundamental plane'' between \mbh, \ls, and $\sigma$ \citep[e.g.,][]{hop07},  
and perform further tests for systematics.  
Due to the failure of NICMOS in Fall 2008, nine objects at $z=0.57$
and three objects at $z=0.36$ were not observed.  Instead, we were
recently allocated time with WFC3 to complete the full sample of
Seyfert-1 galaxies. We will present results for this extended sample
in another paper.  At the same time, increasing the local AGN
comparison sample would be desirable (and indeed an HST proposal for
the eight nearest RM AGNs by Bentz et al. is in the queue).
Understanding slope and scatter of the local relations for active
galaxies is crucial to study their evolution.

\acknowledgments

We thank Chien Peng for his advice on using GALFIT,
his help with the high-z sample and stimulating discussions.
Raphael Gavazzi helped developing the NICMOS reduction
pipeline.  We are grateful to Andrew Benson for providing us with his
measurements in electronic format.  We thank Misty Bentz for help with
the RM AGN data and Phil Marshall for helpful insights.
We thank the anonymous referee for carefully reading
the manuscript and for useful suggestions.
This work is based on data obtained with the Hubble
Space Telescope and the 10 W.M. Keck Telescope and is made possible by
the public archive of the Sloan Digital Sky Survey.  V.N.B. is supported
through a grant from the National Science Foundation (AST-0642621) and
by NASA through grants associated with HST proposals GO 11208, GO 11341,
and GO 11341. T.T. acknowledges support from the
NSF through CAREER award NSF-0642621, from the Sloan Foundation, and
from the Packard Foundation.  
J.W. acknowledges the support provided by NASA through HST grant AR-10986 and
Hubble Fellowship grant HF-0642621 awarded by the Space Telescope Science
Institute,
which is operated by the Association of
Universities for Research in Astronomy, Inc., for NASA, under contract NAS
5-26555.
This research has made use of the
NASA/IPAC Extragalactic Database (NED) which is operated by the Jet
Propulsion Laboratory, California Institute of Technology, under
contract with the National Aeronautics and Space Administration.

{\it Facilities:} \facility{Keck:I (LRIS)}, \facility{HST (NICMOS)}

\appendix
\section{MONTE CARLO SIMULATIONS}
\label{sec:mc}

To probe the reliability of GALFIT to derive the AGN and host-galaxy
properties accurately and to estimate the systematic uncertainties involved in the 
fitting, we ran Monte Carlo simulations of a set of different galaxy models.
A comparable procedure was carried out by \citet{kim08}.

In particular, we used GALFIT to simulate galaxies, consisting of (a) PSF plus spheroid, and
(b) PSF plus spheroid plus disk using a range of typical galaxy properties
of our sample.
In both cases, the total magnitude was set to either 18 or 19 mag.
For case (a), we assumed the effective radius of the spheroid to be
$r_{\rm eff}$ = 4 pix, 6 pix, 8 pix,
and an AGN-to-total luminosity ratio of 0.05, 0.1, 0.2, 0.5, 0.8, 0.9, 0.95.
In case (b), the effective radius of the spheroid was set to
$r_{\rm eff}$ = 3 pix, 4 pix, 6 pix, 8 pix
(to additionally probe the lower limit which can be a problem
when fitting spheroid plus disk).
For an AGN-to-total luminosity ratio of 0.8 (0.5), the spheroid-to-disk ratio was
0.5 (0.2, 0.5), and for an AGN-to-total luminosity ratio
of 0.2 and 0.1, a spheroid-to-disk ratio of 0.1, 0.2, 0.5 was used.

Note that this is a conservative approach, focusing on the parameter space
for which the detection of the spheroid component is most difficult,
i.e.~a small spheroid size, a small spheroid-to-disk luminosity ratio and 
a large AGN-total galaxy luminosity ratio.
The galaxies were simulated with a given PSF and
noise was added based on a typical observed signal-to-noise (SN) ratio
in a Monte Carlo fashion creating 100 artificial images which were
then fitted by GALFIT.
In a first run, the simulated galaxies were fitted with the same PSF
that was used to create the artificial images,
in subsequent runs with different PSFs from our PSF library
to simulate PSF mismatch.

\begin{figure*}[h!]
\includegraphics[scale=0.4]{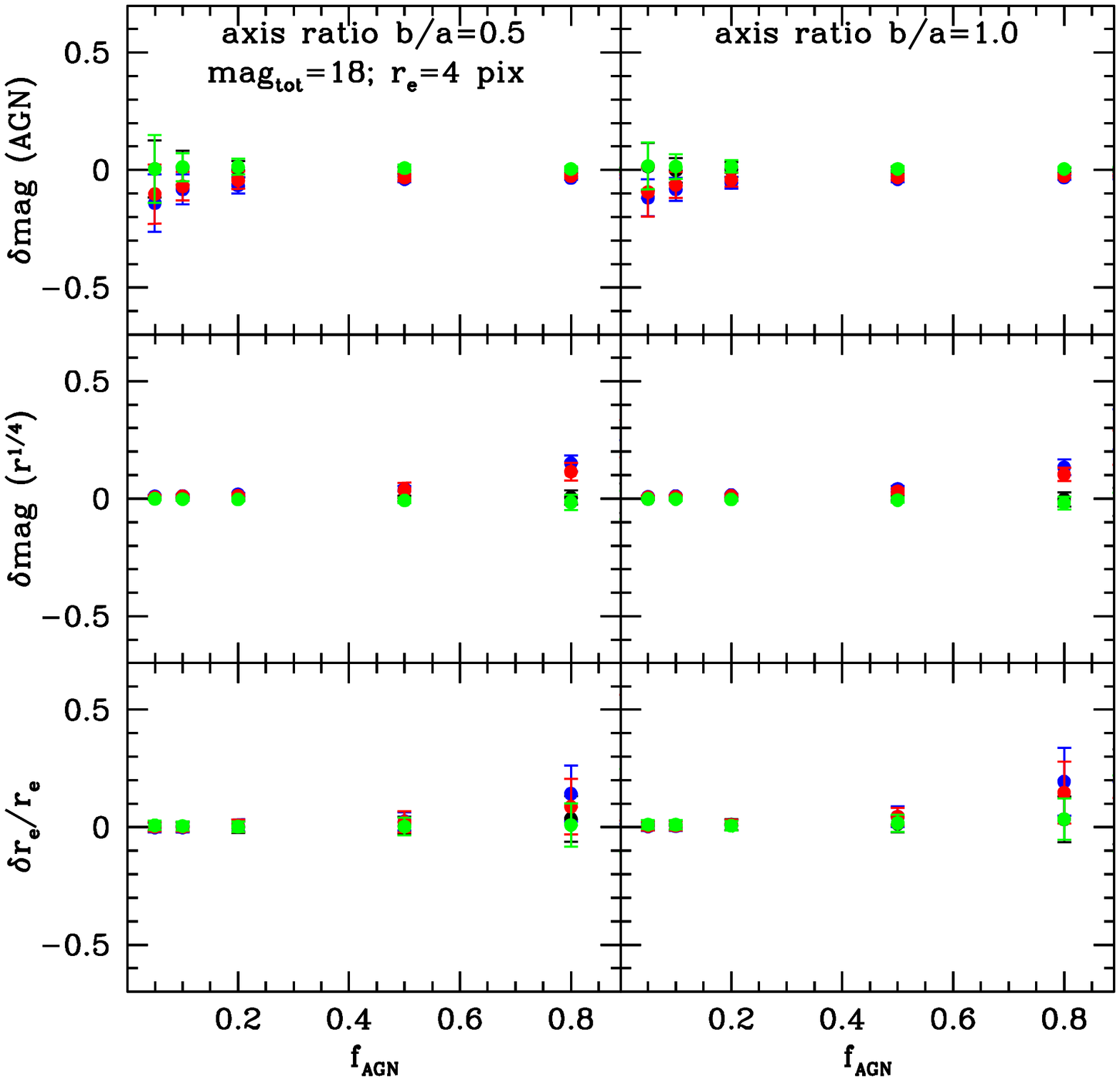}
\includegraphics[scale=0.4]{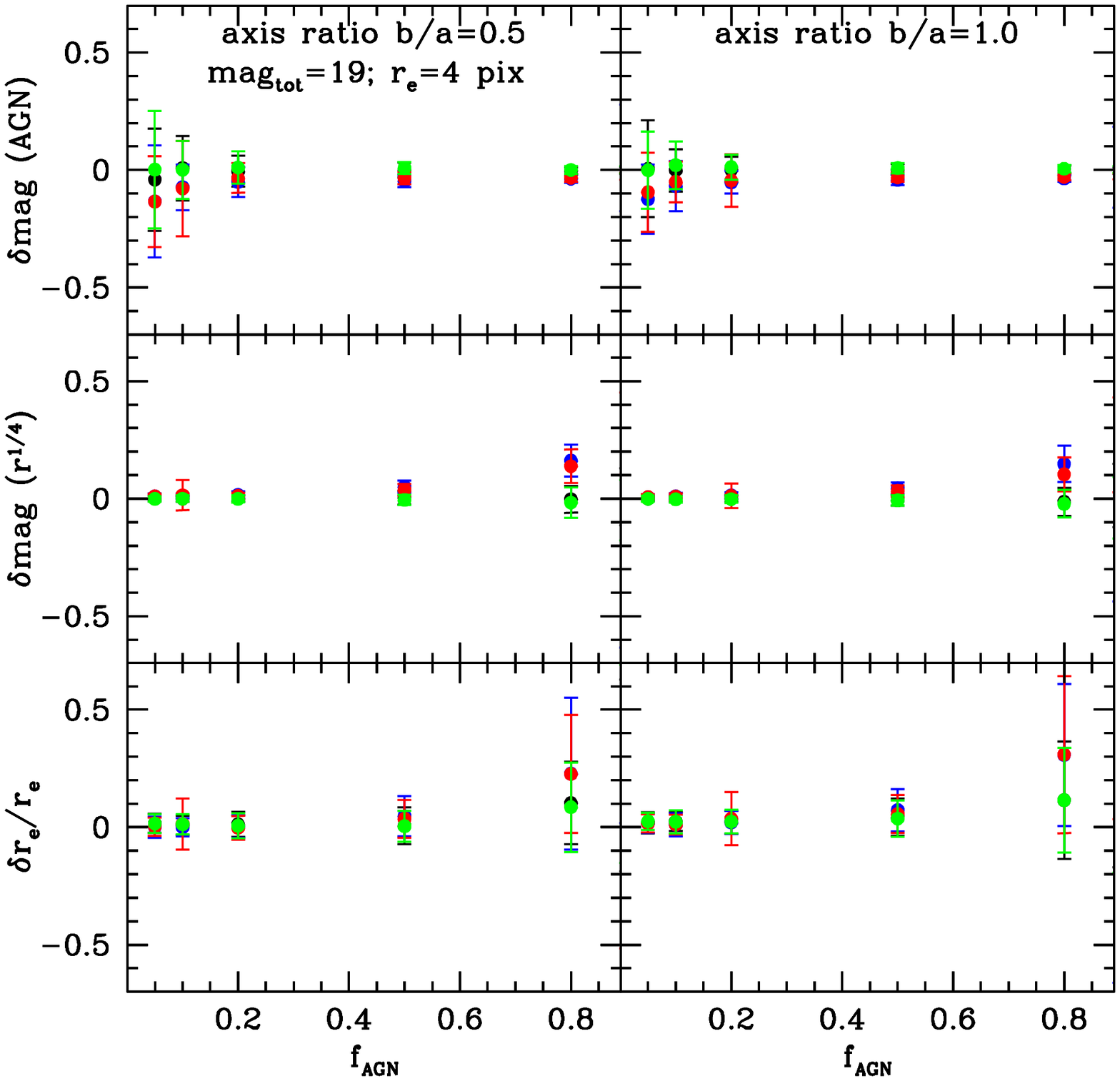}
\caption{Results of GALFIT fits to simulated galaxies,
consisting of PSF plus spheroid, with noise added in a
Monte Carlo fashion, realizing 100 artificial
images for each parameter combination. The difference
between input AGN magnitude and derived AGN magnitude 
is shown (upper panels), the difference between input spheroid magnitude
and derived spheroid magnitude (middle panel), and the difference
between input and derived effective radius of the spheroid (lower panel).
Each data point represents the average plus error of
GALFIT fits to the 100 artificial images. Black data points
correspond to fits where the PSF used to create the artificial
image is identical with the one used for fitting.
The other three colors correspond
to a different PSF used for fitting which was taken from our
PSF library to simulate PSF mismatch.
The left figure shows results for a total host-galaxy magnitude of 18 mag,
the right figure of 19 mag. The left panels within each
figure correspond to an axis ratio of b/a = 0.5, the
right panels to b/a = 0.1. In both figures, the input effective radius
of the spheroid component is set to 4 pixels.}
\label{montecarlobonly}
\end{figure*}

\begin{figure*}[h!]
\includegraphics[scale=0.4]{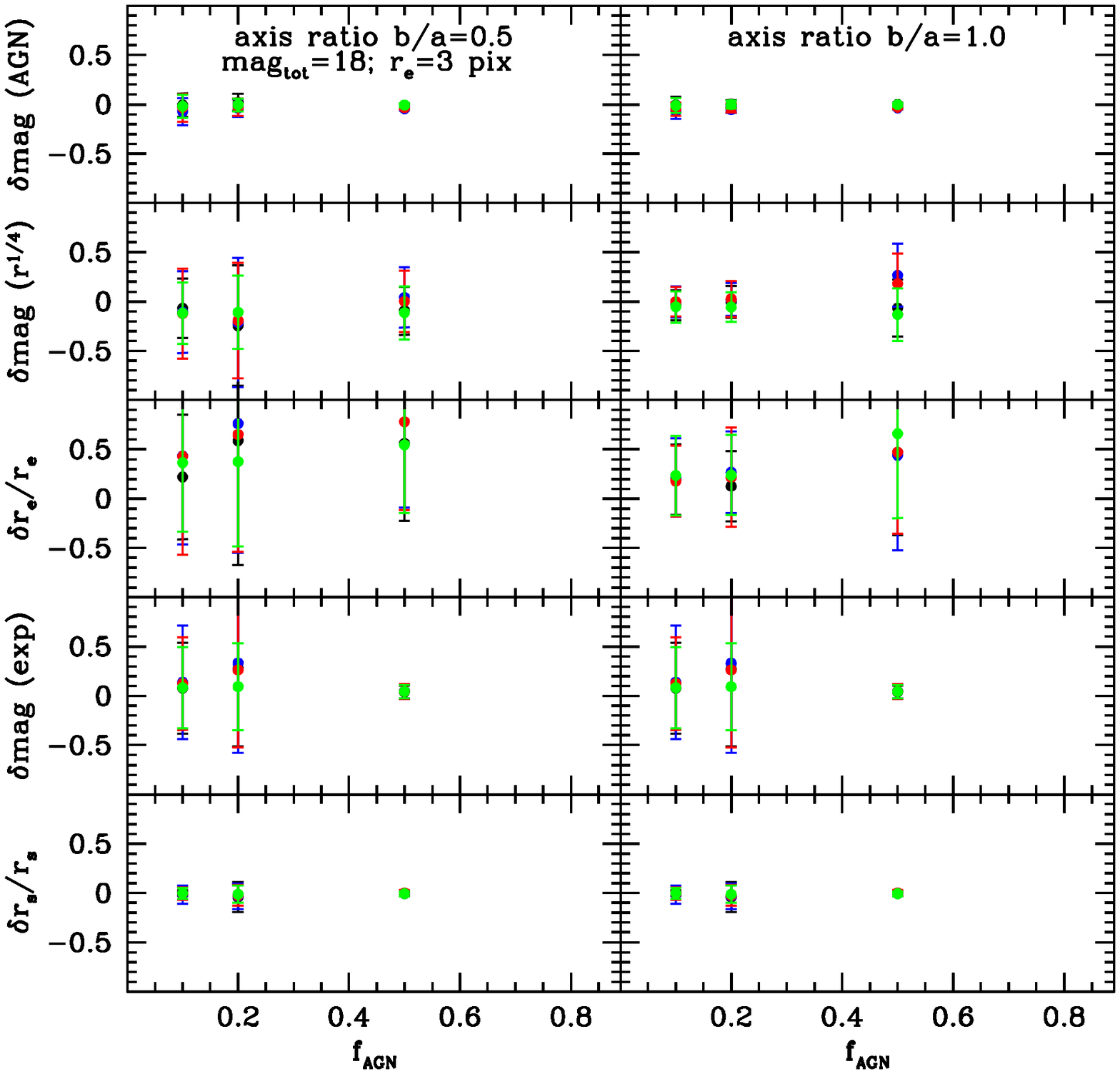}
\includegraphics[scale=0.4]{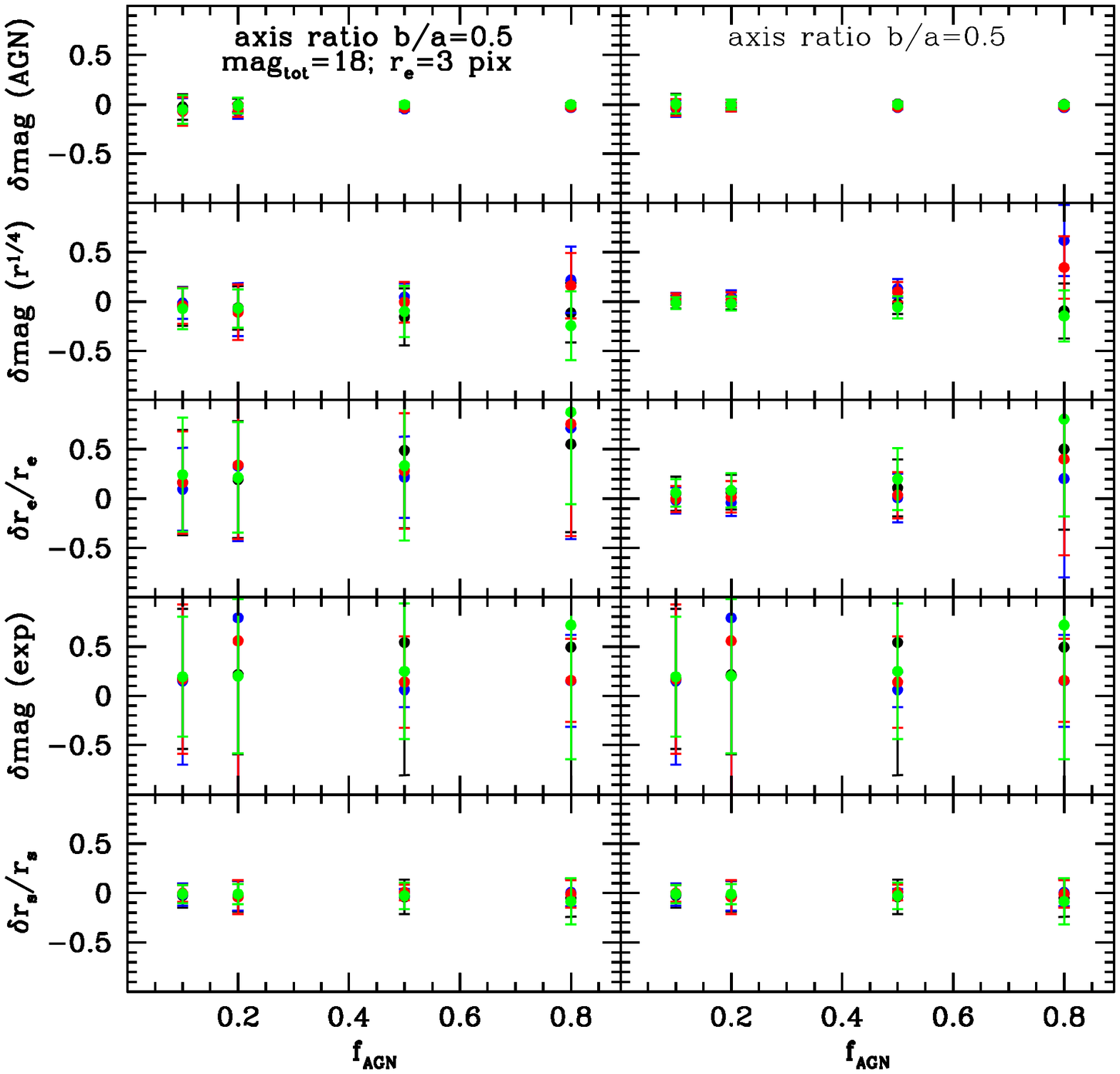}
\caption{The same as in Fig.~\ref{montecarlobonly}
for artificial images consisting of PSF plus spheroid
plus exponential component. The difference between input disk magnitude
and derived disk magnitude and the difference between input disk radius
and derived disk radius is shown additionally in the two lower panels.
In the left figure, the spheroid-to-disk ratio is 0.2
(and thus, the AGN-to-total luminosity $f_{\rm AGN}$ plotted
on the x-axis only assumes values of 0.1, 0.2, 0.5), in the right figure,
the spheroid-to-disk ratio is 0.5 (with $f_{\rm AGN}$ = 0.1, 0.2, 0.5, 0.8;
see text for details).
In both cases, the total host-galaxy magnitude is 18 mag and the effective
radius of the spheroid is set to 3 pixels.
While the PSF magnitude can be retrieved easily to within 0.2 mag,
the difference in spheroid magnitude can be up to 0.5 mag in the
worst case. 
}
\label{montecarlobd}
\end{figure*}

For case (a), 
GALFIT can easily recover the sizes and magnitudes,
even when the spheroid reaches sizes 
close to the minimum size that can be resolved given the PSF (here assumed to be
2.5 pixels). In Fig.~\ref{montecarlobonly}, we show the resulting offsets
for the smallest spheroid ($r_{\rm eff}$ = 4 pix). 
However, more caution needs to be exercised for a three-component fit (PSF, spheroid, and disk),
probed in case (b). In Fig.~\ref{montecarlobd}, we show the resulting offsets
for the smallest spheroid ($r_{\rm eff}$ = 3 pix), i.e.~the most
difficult scenario for retrieving the spheroid parameters accurately. 
The derived spheroid magnitude can differ up to 0.5 mag in
the worst case, while the difference in PSF magnitude is less
than 0.2 mag.
We adopt these values as conservative measures
of our errors, i.e.~0.2 mag for AGN luminosity and 0.5 mag
for spheroid luminosity.

As the estimation of errors is the main purpose of this
analysis, we do not further discuss the results
of these simulations.
The overall trend is the same as in \citet{kim08}, i.e.~the 
scatter in all derived parameters
is largest when the AGN is dominant,
and when $r_{\rm eff}$ is small and difficult to distinguish
from the nucleus or large with low surface brightness.
Spheroid-to-disk-to-AGN decompositions are much more difficult
than spheroid-to-AGN as they involve 6 additional free parameters
(if the spheroid is fitted by a \citet{dev48} profile) and can only
be done if the S/N is high.

\section{CHOICE OF S{\'E}RSIC INDEX}
\label{sec:sersic}

A general profile to fit galaxies is the so-called \citet{ser68} power
law, which is defined as
\begin{equation}
\Sigma (r) = \Sigma_{\rm eff} \exp \left[- \kappa_n \left(\left(\frac{r}{r_{\rm eff}}\right)^{1/n}-1\right)\right] \hspace{0.2cm} ,
\end{equation}
where $\Sigma_{\rm eff}$ is the pixel surface brightness at the
effective radius $r_{\rm eff}$, and $n$ is the S{\'e}rsi{\'c} index.
In this generalized form, an exponential disk profile has $n$ = 1, a
\citet{dev48} profile has $n$ = 4, and a Gaussian has $n$ =
0.5 (which was used in Paper II to fit a bar component).  While
\citet{dev48} profiles are traditionally and widely used to fit
spheroidal components, recent studies show that spheroids can have
S{\'e}rsic indices ranging between 0.5 and 6.
Disk galaxies typically have a bulge component with $n < 4$,
with classical bulges having $n \gtrsim 2$
and pseudobulges having $n \lesssim 2$ \citep{fis08}.
Moreover, there seems to be a relation between 
the S{\'e}rsi{\'c} index and the spheroid 
luminosity or host-galaxy luminosity
\citep[e.g.,][]{kor78,sha89,and94,gra01,pen02,nip08,mac08}.

The objects we are fitting are complex in nature,
in particular due to the presence of the AGN,
a very luminous point source in the center for which a perfectly matching
PSF fit cannot always be achieved.
Thus, we cannot use a fit with the
S{\'e}rsic index $n$ as a free parameter, as it would add yet
another free component to an already difficult fit and increase
degeneracies between PSF, bulge, and disk. 
Such an approach could easily lead to an
unphysical fit, if GALFIT is trying to fit any remaining
PSF mismatch with such a component.
In such a situation, an alternative approach to estimate the 
best fitting S{\'e}rsic exponent is to use a range
of S{\'e}rsic indices, keep them fixed at each step
and then obtain the best $n$ from the resulting 
the $\chi^2$ statistics.
This approach is generally
recommended when attempting galaxy decompositions of faint or
difficult to model galaxies like AGN host galaxies \citep{pen02, kim08}.

To test the systematic uncertainties in derived spheroid and PSF
magnitude depending on the adopted S{\'e}rsic index, we re-ran our
models using $n = 0.5, 1, 2, 3, 4, 5$ to fit the spheroid component.
At each step, we kept $n$ fixed to the chosen value
but allowed all other parameters to vary, including
the disk component.
For those objects that were initially (i.e. when using $n$=4)
fit by a spheroid component only,
we carefully checked the residuals of
the resulting best $n$ fit
for any evidence of an additional disk component.
For only one source (1501) is the quality of the fit
increased significantly by
the addition of a disk component.\footnote{Note, however, that fitting this source
is particularly complicated as it is in the process
of merging with a neighboring galaxy.}
Note that using the best-fitting $n$ instead of
$n$=4 does not in general solve the problem of a a vanishingly small bulge component
for some objects. For 11 of the 16 objects discussed in \S~\ref{ssec:prior},
nothing changes. For five objects, the effective radius of the bulge component
is no longer smaller than the FWHM of the PSF; however, for two different objects,
the bulge component then becomes vanishingly small.

\begin{figure*}[h!]
\includegraphics[scale=0.4]{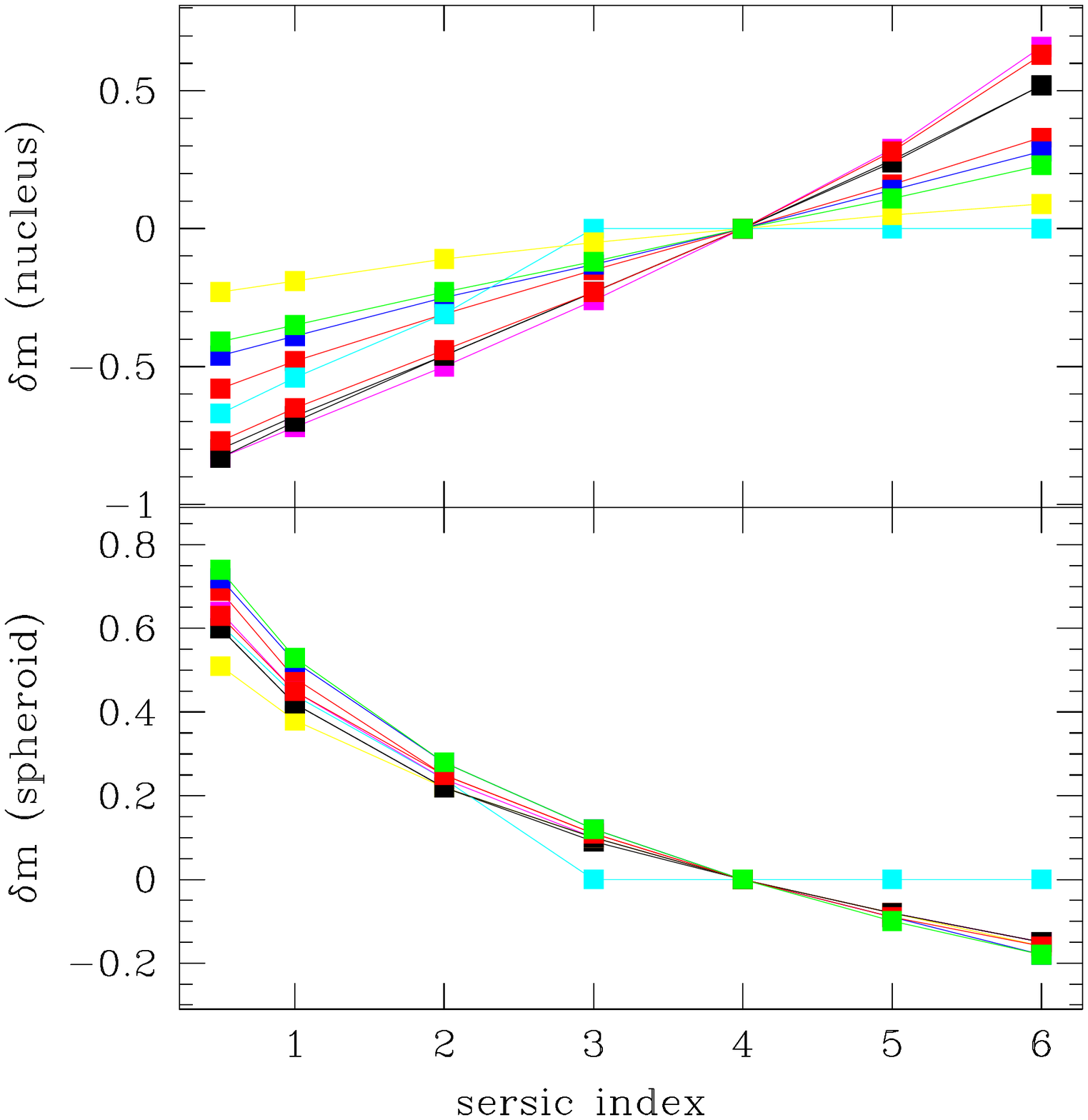}
\includegraphics[scale=0.4]{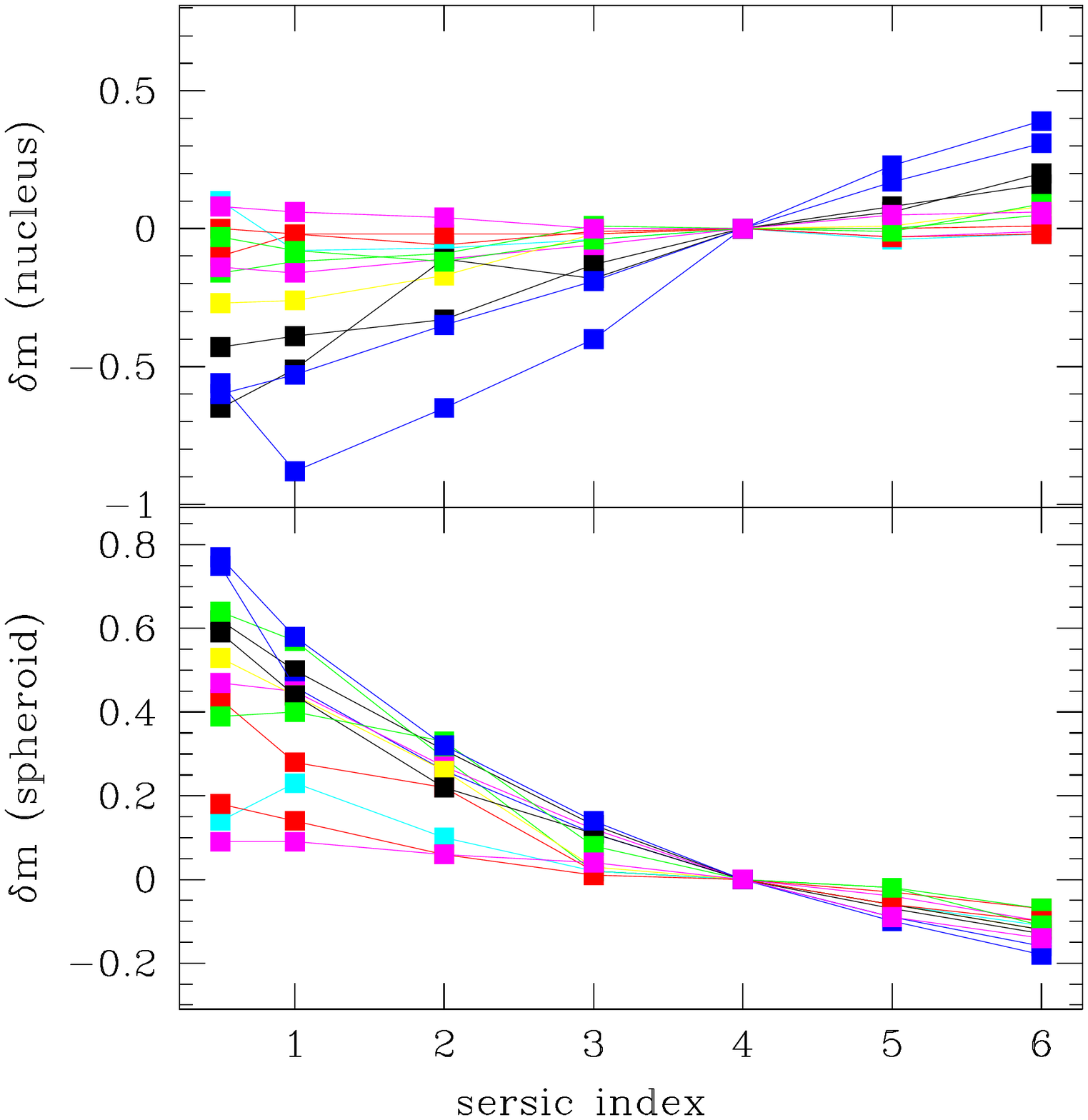}
\caption{Systematic effects in derived magnitudes of spheroidal
and AGN component due to the adopted spheroid profile (S{\'e}rsic index 0.5, 1, 2, 3, 4, 5,
or 6). {\bf Left panel:} All objects for which the host
galaxy was fitted by a spheroid component
only, with each object corresponding to a given color. {\bf Right panel:}
Same as the in the left, for objects for which the host galaxy was fitted with a spheroid plus disk component.
See text for details. [{\sl See the electronic edition of the Journal
for a color version of this figure.}]}
\label{sersic}
\end{figure*}

The results are shown in Fig.~\ref{sersic}, separating objects for which
the host galaxy was fitted by a spheroid component only (left panel) and those
for which the host was fitted by a spheroid plus disk component (right panel).
While the overall trend is the same, fitting the host galaxy by two
components results in a larger scatter because the disk
magnitude can also vary. 
The results can be summarized as follows: 
Decreasing $n$ from 4 to 3 (2, 1, 0.5) decreases the spheroid
luminosity -- on average by 0.08 (0.23, 0.41, 0.54) mag -- 
and increases the nuclear luminosity -- on average by -0.07 (-0.24, -0.37, -0.4) mag. 
Increasing $n$
from 4 to 5 (6) on the other hand increases the spheroid luminosity on
average by -0.07 (-0.13) mag and decreases the nuclear luminosity on average by
0.1 (0.22) mag.
For all but the most extreme indices, potential
systematics related to the choice of S{\'e}rsic index are small
compared to the adopted uncertainty on the spheroid luminosity (0.5 mag) and on
\mbh~(0.4 dex).

Another approach is to calculate the best S{\'e}rsic index we would expect based on
the measured host-galaxy magnitude using the relation in
\citet{nip08}, derived from surface-photometry of
ACS images of the well-defined Virgo cluster sample 
\citep{fer06,gal08}
\begin{equation}
\log n = (0.27 \pm 0.02) \log \left(\frac{L_{\rm sph, B}}{L_{\odot}} - 9.27\right)+0.4 \pm 0.02
\end{equation}
For the 23 objects studied here, we estimate a S{\'e}rsic index
ranging from $\sim$4.0 to 5.9, on average 4.9 $\pm$ 0.5.  For the
sample studied in Paper II, the S{\'e}rsic index ranges from $\sim$4
to 6.8, with an average of 5.5 $\pm$ 0.6.  
Within the errors, these values are in agreement with estimates
using the relation between S{\'e}rsic index and bulge B-band magnitude
for a local sample from \citet{gra01} (their Figure 14, middle panel).
Note that in both cases, the estimated S{\'e}rsic index remains the same within
the errors when using the host-galaxy luminosity (for n-L relation
from \citealt{nip08}) or bulge luminosity (for n-L relation from \citealt{gra08})
as derived from the best fit with a free S{\'e}rsic index 
instead of the one derived from $n$=4.
As the relation between
S{\'e}rsic index and host-galaxy magnitude has its own uncertainties
and scatter, and as our average value is close
to 4, we adopted the simpler solution of fixing
$n$ to 4 for all objects as our default choice. This also allows a better
comparison with other AGN host-galaxy studies.

To ultimately probe the potential systematics related
to the choice of S{\'e}rsic index, we performed the same analysis
as for $n$=4, but this time using the best $n$ derived from the procedure described above
(i.e. as chosen based on the $\chi^2$ statistics when 
performing a variety of fits with $n$ fixed to 0.5, 1, 2, 3, 4, 5, 6)
both for the intermediate-z and the local sample.
The same approach was followed for the local sample of RM AGNs.
None of the results stated in the paper change: The resulting
fits, offsets, and predicted evolutionary trends remain the same within the errors.
More precisely, for the evolution in \mbh (\mbh/\ls$\propto$$(1+z)^{\beta}$),
including selection effects, we 
obtain $\beta = 1.3 \pm 0.2$ (instead of $\beta = 1.4 \pm 0.2$ for $n=4$) for the full sample with an intrinsic scatter
$<$0.7 dex at 95\% CL (0.4$\pm$0.1 dex for a uniform prior on $\sigma_{\rm int}$)
and $\beta = 3.4 \pm 1.2$ (instead of $\beta = 2.8 \pm 1.2$ for $n=4$) for the intermediate sample alone,
adopting again the prior by \citet{gue09}.

\section{SURFACE PHOTOMETRY OF RM AGNs}
\label{sec:rm}
For an homogeneous treatment of all data,
we performed an independent analysis of the HST
archival images presented in \citet{ben09b},
using the same approach as for our distant Seyfert galaxies
(\S~\ref{sec:surface}). Details of the observations can be found
in \citet{ben09b}. 

We disregarded the five objects observed with WFPC2/PC
due to the low quality of the data, the PSF mismatch
when using a synthetic PSF created by TinyTim and the lack
of stellar PSFs on the images. These problems made it difficult
to achieve satisfactory fits. The spheroid radius
found with GALFIT was either in the lower limit of
2.5 pixels (=FWHM of PSF) -- probably because it was
fitting a PSF mismatch -- or was unphysically large.
Thus, we here focus on the ACS/HRC data alone.

From the remaining 30 objects imaged with ACS/HRC,
we first excluded all NGC objects (8/30) which
are nearby and extended and for which the field-of-view is
too small to measure the sky background. Also, they are often affected by dust lanes.
The latter is also the case for IC\,4239A (plus an unreliable BH mass). 
For the same reasons, these objects
were also excluded in the further analysis by \citet{ben09a}.
We decided to additionally exclude 
Fairall 9 due to a dust lane crossing the spheroid
and PG0953+414 for which no reasonable fit could be achieved.
Thus, our final robust sample consists of 19 objects.

We used the pipeline-processed data and combined them using multidrizzle,
to remove cosmic rays and defects and correct for distortion.
(Note that multidrizzle takes into account the saturated pixels
of the longer exposures and combines the images accordingly.)
As the data were not dithered, no improvement of sampling
was achieved and the final scale is 0.025 arcsec/pixel (pixfrac=0.9).
For these ACS/HRC data imaged in the F550M filter,
the PSF created by TinyTim is not as good
a match as it is for the NICMOS images.
We therefore additionally created a PSF from a star
observed in one of the images (Mrk 110) and performed
extensive tests to compare their quality. As the TinyTim PSF
typically gave a bad fit in the core, but the stellar PSF
had too low S/N in the wings, we decided to combine
both PSFs (the synthetic PSF for the wings, the stellar PSF in center out to $r$ = 2 $\times$ FWHM),
which significantly improved the quality of the fits.
This PSF enabled us to fit the AGN with only one PSF
without the need of corrections of PSF mismatch
(e.g.~by using an additional PSF as done by \citet{ben09b}).
We used the same criteria as for our
distant Seyfert sample to decide whether we need to fit an additional
disk component (see \S~\ref{sec:surface}).
For four objects, \citet{ben09b} 
fitted both a spheroidal and disk component, while we decided that fitting a spheroidal component alone
is sufficient.
One object has a saturated PSF (PG1226+023) and we masked out the saturated
center to fit the PSF to the wings only.
For three objects 
(Ark~120, Mrk~279, and PG~1211+143),
we out-masked the very center of the PSF
and fitted the PSF to the wings only due to remaining PSF mismatch.

\begin{figure*}[h!]
\includegraphics[scale=0.25]{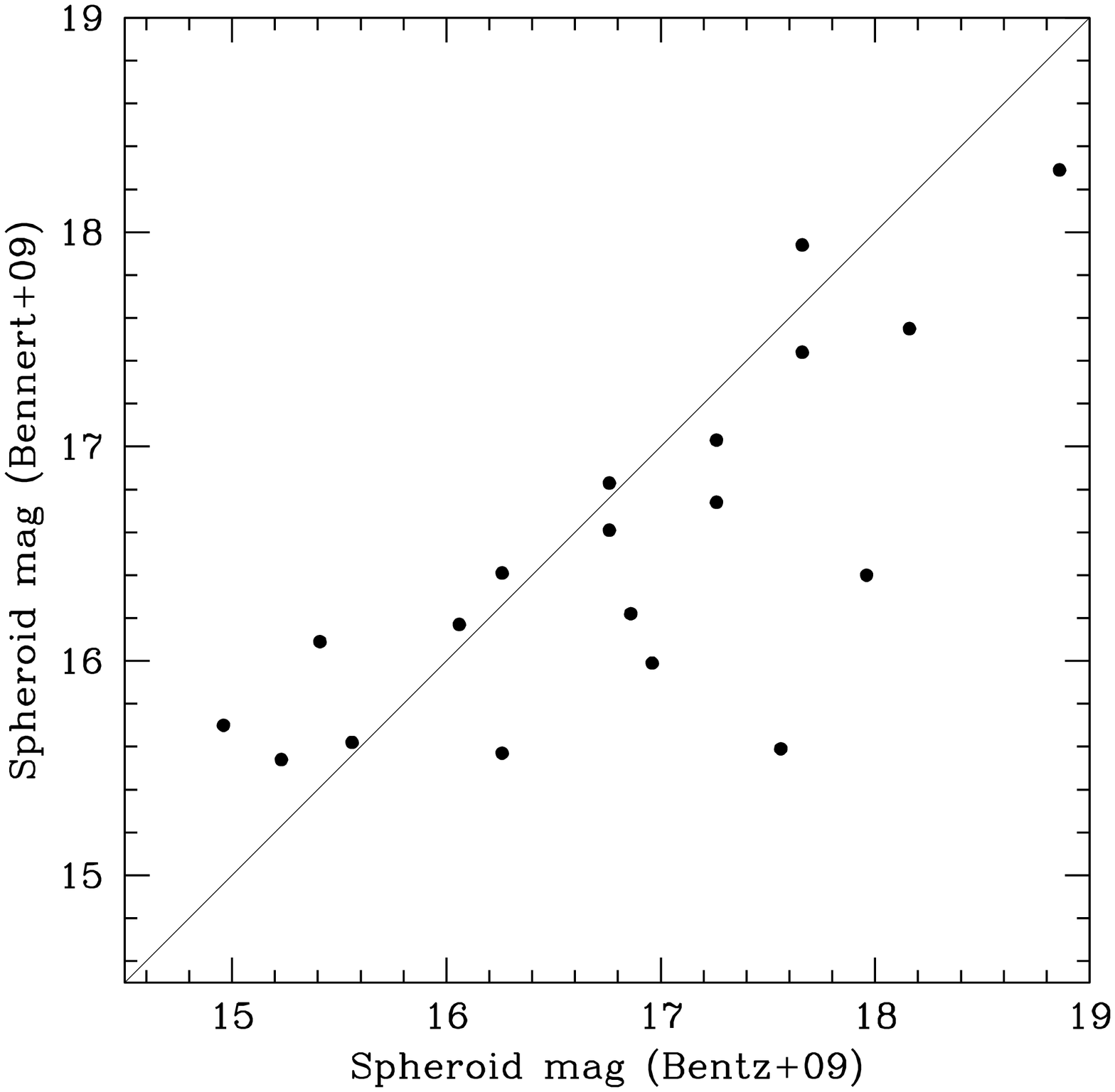}
\includegraphics[scale=0.25]{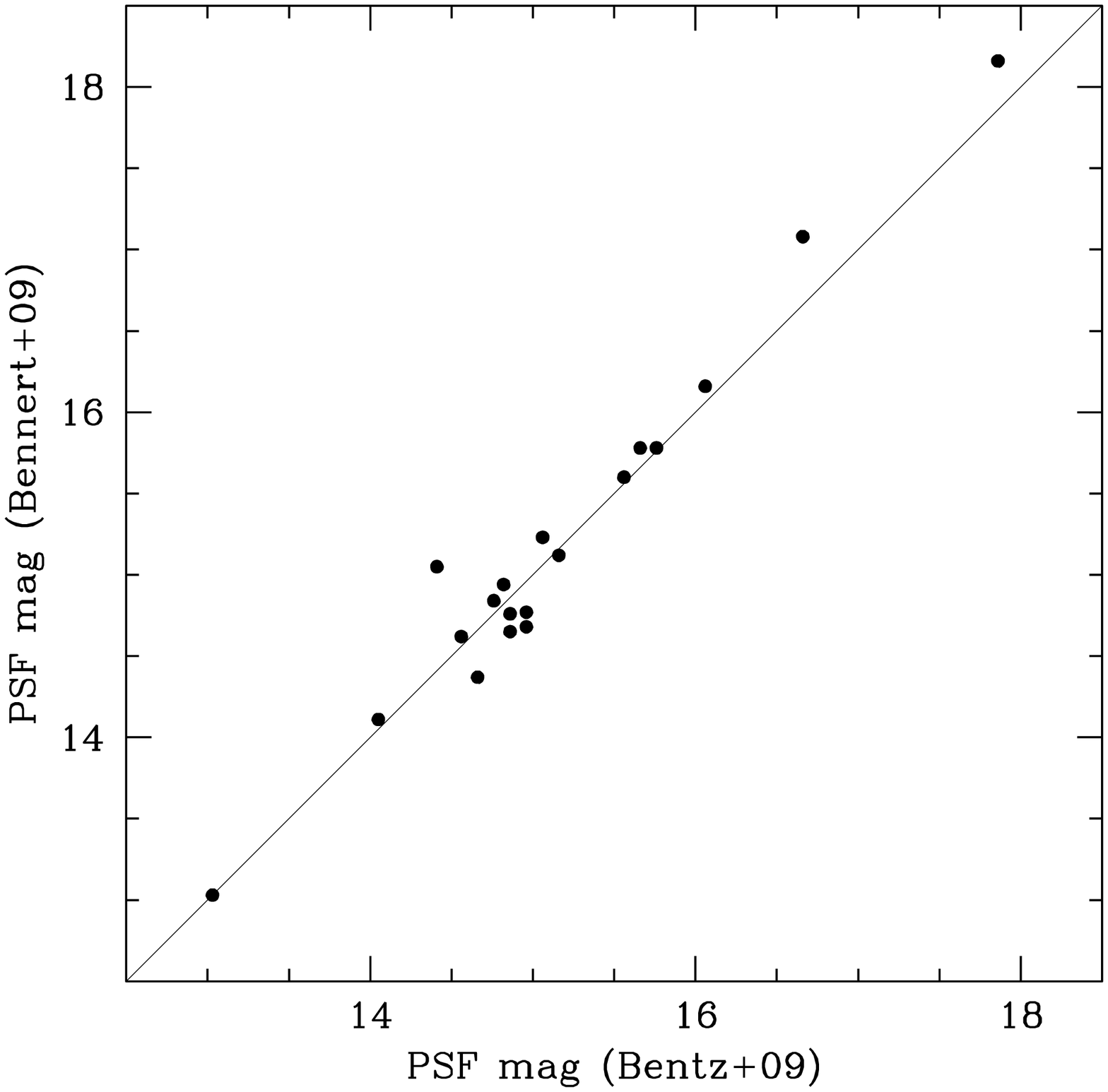}
\includegraphics[scale=0.25]{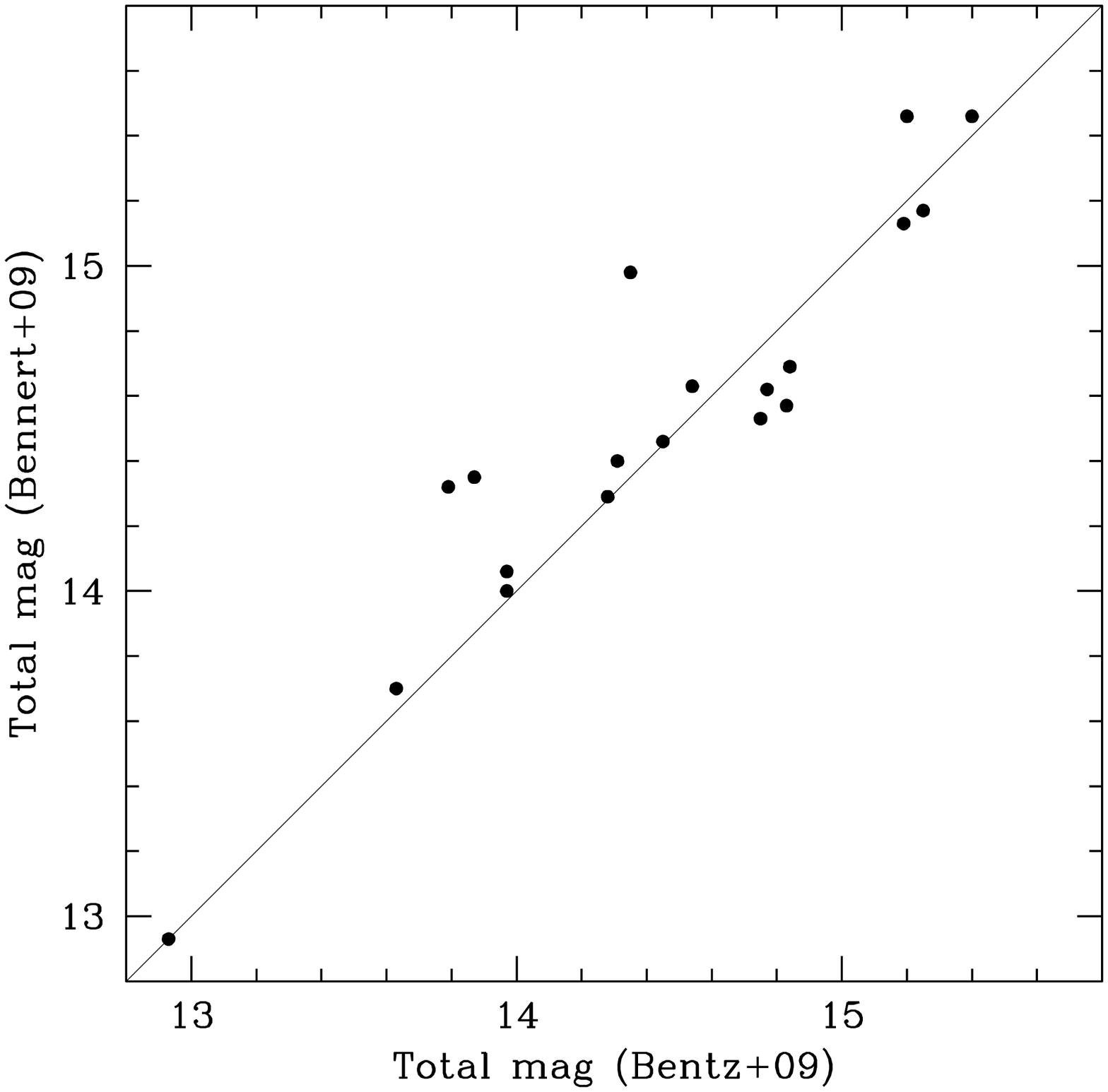}
\caption{Difference between the results of the surface-brightness
fitting of \citet{ben09b} and our work here, for PSF magnitude
({\bf left panel}), spheroid magnitude ({\bf middle panel}),
and total magnitude ({\bf right panel}).}
\label{compare}
\end{figure*}

We compare the results in Fig.~\ref{compare}.
For this comparison, we add the different PSF components and the 
different spheroid components  of \citet{ben09b} to a ``total'' PSF magnitude
and ``total'' spheroid magnitude, respectively. 
While the PSF and total magnitudes generally
agree well, the spheroid magnitudes we derive are often brighter than
in \citet{ben09b}, especially in those cases where we fit a spheroid
component only and not spheroid+disk as in \citet{ben09b} (4 objects).

As for our intermediate redshift sample (see Appendix~\ref{sec:sersic}),
we also calculated the best S{\'e}rsic index we would expect based on
the measured host-galaxy magnitude using the relation in
\citet{nip08}. The average value of $n$=5.2 $\pm$ 1.6 is in agreement
within the errors with the average $n$ derived
when using the relation between S{\'e}rsic index 
and bulge B-band magnitude from \citet{gra01}.
It also agrees well with the average $n$ estimated
for the intermediate-z sample, 
with a larger scatter due to the larger spread in
luminosities.
We carefully checked whether when using 
the best-fitting S{\'e}rsic index,
there is the need of adding a disk component
for those objects for which the host galaxy was originally fitted
by a $n$=4 component only; we do not find such evidence in any of the objects.

\clearpage

\begin{deluxetable}{lccccc}
\tabletypesize{\scriptsize}
\tablecolumns{6}
\tablewidth{0pc}
\tablecaption{Sample Properties}
\tablehead{
\colhead{Name} & \colhead{$z$} & \colhead{$D_{\rm L}$} & \colhead{RA (J2000)} & \colhead{DEC (J2000)} 
& \colhead{$i^\prime$}\\
& & Mpc & & & mag \\
\colhead{(1)} & \colhead{(2)} & \colhead{(3)}  & \colhead{(4)} & \colhead{(5)} & \colhead{(6)}}
\startdata
0107 (S11) & 0.3558  &  1892.9 & 01 07 15.97	& --08 34 29.4   & 18.47 \\
0804 (SS1)  & 0.3566 &  1897.9 & 08 04 27.99  & +52 23 06.2    & 18.55 \\
0934 (SS2)  & 0.3672 &  1964.1 & 09 34 55.60  & +05 14 09.1    & 18.82 \\  
1007 (SS5)  & 0.3733 &  2002.5 & 10 07 06.26  & +08 42 28.4    & 18.69 \\
1015 (S31)  & 0.3505 &  1860.0 & 10 15 27.26   & +62 59 11.5	& 18.15 \\
1021 (SS6)  & 0.3584 &  1909.1 & 10 21 03.58  & +30 47 55.9    & 18.92 \\
1043 (SS7) & 0.3618  &  1930.3 & 10 43 31.50   & --01 07 32.8	& 18.82 \\
1046 (SS8)  & 0.3656 &  1954.1 & 10 46 10.60  & +03 50 31.2    & 18.45 \\
1258 (SS9)  & 0.3701 &  1982.3 & 12 58 38.71  & +45 55 15.5    & 18.56 \\
1334 (SS10) & 0.3658 &  1955.4 & 13 34 14.84  & +11 42 21.5    & 17.83 \\
1352 (SS11) & 0.3732 &  2001.8 & 13 52 26.90  & +39 24 26.8    & 18.39 \\
1501 (SS12) & 0.3625 &  1934.7 & 15 01 16.83  & +53 31 02.4    & 17.80 \\
1505 (SS13) & 0.3745 &  2010.0 & 15 05 41.79  & +49 35 20.0    & 18.73 \\
1611 (S28)  & 0.3679 &  1968.5 & 16 11 56.30   & +45 16 11.0	& 18.63 \\
2115 (SS14)& 0.3706  &  1985.5 & 21 15 31.68   & --07 26 27.5	& 19.24 \\
2158 (S29) & 0.3575  &  1903.5 & 21 58 41.93	& --01 15 00.3   & 18.95 \\
2340 (SS18) & 0.3582 &  1907.9 & 23 40 50.52  & +01 06 35.5    & 18.50 \\
\hline
0155 (W11) & 0.5634  &  3270.9 & 01 55 16.18	& --09 45 56.0   & 20.09  \\
0342 (W22) & 0.5648  &  3280.8 & 03 42 29.70	& --05 23 19.5   & 18.70  \\
1439 (W12)  & 0.5623 &  3263.2 & 14 39 55.11   &  +35 53 05.4	& 19.02  \\
1500 (W20)  & 0.5753 &  3354.7 & 15 00 14.81   &  +32 29 40.4	& 19.60  \\  
1526 (W16) & 0.5782  &  3375.2 & 15 26 54.93	& --00 32 43.3   & 19.99  \\	    
1632 (W8)   & 0.5703 &  3319.4 & 16 32 52.42   &  +26 37 49.1	& 18.70  \\
\enddata
\tablecomments{
Col. (1): Target ID (RA: hhmm). In brackets, the name used in other publications.
Col. (2): Redshift from SDSS-DR7.
Col. (3): Luminosity distance in Mpc, based on redshift and the adapted cosmology.
Col. (4): Right Ascension. 
Col. (5): Declination. 
Col. (6): Extinction-corrected
$i^\prime$ AB magnitude from SDSS-DR7 photometry (``modelMag\_i'').}
\label{data}
\end{deluxetable}

\begin{deluxetable}{lcccccccccc}
\rotate
\tabletypesize{\scriptsize}
\tablecolumns{11}
\tablewidth{0pc}
\tablecaption{Results from Imaging of Distant Seyfert Sample}
\tablehead{
\colhead{Name} & \colhead{Total}  & \colhead{Host} & \colhead{Spheroid} & $\log L_{\rm host, V}/L_{\odot}$  & $\log L_{\rm sph, V}/L_{\odot}$ & $R_e$ & 
\colhead{$\lambda L_{\rm 5100}$} & \colhead{$f_{\rm nuc}$} & \colhead{$\log$ \mbh/$M_{\odot}$}
& \colhead{comp.}\\
& mag & mag & mag & & & kpc &  10$^{44}$ erg\,s$^{-1}$ & & \\
\colhead{(1)} & \colhead{(2)} & \colhead{(3)}  & \colhead{(4)} & \colhead{(5)} & \colhead{(6)} & \colhead{(7)} & \colhead{(8)} & \colhead{(9)} & \colhead{(10)}& \colhead{(11)}}
\startdata
0059+1538 (S09; M)  & 18.24 & 18.51 & 19.08$\pm$0.50 & 10.91 & 10.68	 & 3.24 & 0.71 & 0.22 & 8.13& 3\\
0101--0945 (S10) & 18.03 & 18.37 & 19.32$\pm$0.58 & 10.95 & 10.28(10.57) & 0.49 & 1.03 & 0.27 & 8.25& 3\\
0213+1347 (S12)  & 18.20 & 18.56 & 21.23$\pm$0.60 & 10.90 &  9.56(9.83)  & 0.54 & 0.97 & 0.28 & 8.67& 3\\
1105+0312 (S21; M)  & 17.49 & 17.94 & 18.99$\pm$0.58 & 11.13 & 10.39(10.71) & 0.51 & 2.15 & 0.34 & 8.79& 3\\
1119+0056 (S16)  & 19.16 & 19.87 & 22.28$\pm$0.50 & 10.41 &  9.45	 & 0.76 & 0.73 & 0.48 & 8.27& 3\\
1400--0108 (S23) & 18.02 & 18.39 & 20.88$\pm$0.55 & 10.94 &  9.72(9.95)  & 0.57 & 1.11 & 0.29 & 8.70& 4\\
1400+0047 (S24)  & 18.09 & 18.22 & 18.61$\pm$0.50 & 11.05 & 10.89	 & 12.65& 0.44 & 0.11 & 8.33& 3\\
1529+5928 (S26)  & 18.88 & 19.22 & 20.07$\pm$0.50 & 10.67 & 10.33	 & 0.75 & 0.52 & 0.27 & 8.02& 3\\
1536+5414 (S27; M/I)  & 18.53 & 19.00 & 19.48$\pm$0.50 & 10.75 & 10.56	 & 4.78 & 0.95 & 0.36 & 8.10& 3\\
1539+0323 (S01; M)  & 18.54 & 18.91 & 19.97$\pm$0.50 & 10.77 & 10.34	 & 5.30 & 0.72 & 0.29 & 8.20& 4\\
1611+5131 (S02)  & 19.04 & 19.32 & 19.87$\pm$0.50 & 10.58 & 10.36	 & 2.63 & 0.34 & 0.22 & 7.98& 3\\
1732+6117 (S03; M)  & 17.97 & 18.50 & 20.25$\pm$0.53 & 10.92 &  9.97(10.22) & 0.50 & 1.64 & 0.39 & 8.28& 4\\
2102--0646 (S04) & 18.12 & 18.61 & 20.18$\pm$0.50 & 10.88 & 10.25	 & 0.96 & 1.33 & 0.36 & 8.44& 4\\
2104--0712 (S05) & 18.00 & 18.68 & 20.51$\pm$0.50 & 10.83 & 10.10	 & 1.03 & 1.85 & 0.47 & 8.74& 4\\
2120--0641 (S06) & 18.48 & 18.70 & 20.62$\pm$0.50 & 10.88 & 10.11	 & 1.01 & 0.51 & 0.18 & 8.16& 4\\
2309+0000 (S07; M/I)  & 17.82 & 18.48 & 20.39$\pm$0.50 & 10.91 & 10.15	 & 1.01 & 2.10 & 0.45 & 8.53& 3\\
2359--0936 (S08) & 18.33 & 18.89 & 21.77$\pm$0.50 & 10.77 &  9.62	 & 1.23 & 1.22 & 0.40 & 8.10& 4\\
\hline					     	     
0107--0834 (S11) & 17.85 & 18.01 & 18.84$\pm$0.50  & 10.84 & 10.51	  & 0.59 & 0.52 & 0.14 & 8.00 & 3\\ 
0804+5223 (SS1)  & 17.89 & 18.01 & 19.34$\pm$0.58  & 10.84 & 10.04(10.31) & 0.47 & 0.39 & 0.11 & 7.75 & 3\\ 
0934+0514 (SS2; M/I)  & 18.38 & 18.53 & 18.53$\pm$0.50  & 10.67 & 10.67	  & 2.55 & 0.33 & 0.13 & 7.72 & 2\\ 
1007+0842 (SS5)  & 18.34 & 18.80 & 19.69$\pm$0.65  & 10.58 &  9.89(10.22) & 0.49 & 0.93 & 0.34 & 7.66 & 3\\ 
1015+6259 (S31)  & 17.83 & 17.91 & 18.67$\pm$0.50  & 10.86 & 10.56	  & 1.07 & 0.29 & 0.08 & 7.94 & 3\\ 
1021+3047 (SS6; M?)  & 18.85 & 19.14 & 20.29$\pm$0.70  & 10.40 &  9.60(9.94)  & 0.48 & 0.37 & 0.24 & 7.47 & 3\\ 
1043--0107 (SS7) & 18.31 & 18.45 & 19.31$\pm$0.63  & 10.68 & 10.02(10.34) & 0.51 & 0.31 & 0.12 & 7.53 & 3\\ 
1046+0350 (SS8)  & 17.89 & 18.04 & 19.67$\pm$0.55  & 10.86 &  9.95(10.20) & 0.48 & 0.51 & 0.13 & 7.89 & 3\\ 
1258+4555 (SS9)  & 18.04 & 18.37 & 18.37$\pm$0.50  & 10.74 & 10.74	  & 1.62 & 0.93 & 0.26 & 8.05 & 2\\ 
1334+1142 (SS10) & 17.58 & 18.19 & 18.68$\pm$0.65  & 10.80 & 10.25(10.60) & 0.48 & 2.26 & 0.43 & 7.94 & 3\\ 
1352+3924 (SS11) & 18.13 & 18.31 & 19.44$\pm$0.60  & 10.77 & 10.02(10.32) & 0.49 & 0.51 & 0.15 & 8.11 & 3\\ 
1501+5331 (SS12; M) & 17.38 & 18.19 & 18.19$\pm$0.58  & 10.79 & 10.69(10.79) & 0.48 & 3.24 & 0.52 & 8.15 & 2\\ 
1505+4935 (SS13) & 18.40 & 18.92 & 18.92$\pm$0.50  & 10.53 & 10.53	  & 1.09 & 0.98 & 0.38 & 7.63 & 2\\ 
1611+4516 (S28)  & 18.08 & 18.11 & 18.86$\pm$0.50  & 10.84 & 10.54	  & 0.94 & 0.11 & 0.03 & 7.90 & 3\\ 
2115--0726 (SS14)& 18.97 & 19.20 & 19.20$\pm$0.50  & 10.41 & 10.41	  & 1.59 & 0.29 & 0.19 & 7.64 & 2\\ 
2158--0115 (S29; M?) & 18.36 & 18.48 & 19.18$\pm$0.65  & 10.66 & 10.05(10.38) & 0.48 & 0.25 & 0.10 & 7.94 & 3\\ 
2340+0105 (SS18) & 18.41 & 18.79 & 20.20$\pm$0.65  & 10.53 &  9.66(9.97)  & 0.48 & 0.70 & 0.30 & 7.51 & 3\\ 
\hline	  				
0155--0945 (W11; M/I?) & 19.64 & 19.82 & 19.82$\pm$0.50  & 10.64 & 10.64	  & 2.17 & 0.31 & 0.15 & 8.00 & 2\\ 
0342--0523 (W22; M/I?) & 18.05 & 18.53 & 18.53$\pm$0.50  & 11.16 & 11.16	  & 7.34 & 3.17 & 0.36 & 8.65 & 2\\ 
1439+3553 (W12)  & 18.53 & 18.96 & 19.21$\pm$0.65  & 10.98 & 10.54(10.88) & 0.62 & 1.87 & 0.33 & 8.72 & 3\\ 
1500+3229 (W20)  & 19.00 & 19.16 & 19.16$\pm$0.50  & 10.93 & 10.93	  & 3.08 & 0.51 & 0.13 & 8.52 & 2\\ 
1526--0032 (W16; M) & 19.33 & 19.58 & 19.58$\pm$0.50  & 10.76 & 10.76	  & 1.33 & 0.60 & 0.21 & 7.59 & 2\\ 
1632+2637 (W8)   & 18.48 & 19.08 & 19.08$\pm$0.50  & 10.95 & 10.95	  & 1.52 & 2.59 & 0.42 & 8.73 & 2\\ 
\enddata
\tablecomments{
Col. (1): Target ID (RA: hhmm). In brackets, the name used in other publications. 
Additionally, M/I marks objects that are merging/interacting (see Fig.~\ref{nicmos} and Paper II).
All S* and SS* objects are at $z \simeq 0.36$, all W* objects are at $z \simeq 0.57$.
The first 17 objects were observed with ACS/F775W and 
are already included in \citet{tre07}, but are listed here
again due to a small error in extinction correction ($<$0.15 mag); also the luminosity in V 
was not included in \citet{tre07}. For those objects with
upper limits in Paper II, we here estimate the spheroid luminosity using priors (\S\ref{ssec:prior}).
Col. (2): Total extinction-corrected F110W AB magnitude (SS* and W* objects)
or F775W AB magnitude (S* objects); including PSF (with an uncertainty of 0.2 mag).
Col. (3): Total host-galaxy extinction-corrected F110W (F775W) AB magnitude (with an uncertainty of 0.1 mag). 
Col. (4): Spheroid extinction-corrected F110W (F775W) AB magnitude (with an uncertainty of 0.5 mag). 
Col. (5): Logarithm of total host-galaxy luminosity in rest-frame V (solar units), not corrected for evolution.
Col. (6): Logarithm of spheroid luminosity in rest-frame V (solar units), not corrected for evolution.
For those objects, for which the fitting procedure ran into the lower limit of the spheroid effective
radius, we give the corresponding value as upper limit in brackets.
Col. (7): Spheroid effective radius (in kpc; semi-major axis).
Col. (8): Nuclear rest-frame luminosity at 5100$\AA$ (in 10$^{44}$ erg\,s$^{-1}$) (uncertainty of 20\%).
Col. (9): Nuclear light fraction in F110W (F775W) (uncertainty of 20\%).
Col. (10): Logarithm of BH mass (solar units) (uncertainty of 0.4 dex).
For those objects, for which the fitting procedure ran into the lower limit of the spheroid effective
radius, we give the corresponding value as upper limit in brackets.
Col. (11): Number of components fitted (2=PSF+spheroid; 3=PSF+spheroid+disk; 4=PSF+spheroid+disk+bar).
}
\label{results}
\end{deluxetable}

\begin{deluxetable}{lcccccccccc}
\tabletypesize{\scriptsize}
\tablecolumns{11}
\tablewidth{0pc}
\tablecaption{Results from Imaging of Local Comparison Sample}
\tablehead{
\colhead{Name} & \colhead{$z$} & \colhead{$D_{\rm L}$} & \colhead{Total} & \colhead{Host} & \colhead{Spheroid} & $\log L_{\rm host, V}/L_{\odot}$ & $\log L_{\rm sph, V}/L_{\odot}$ & $R_e$ & \colhead{$\log$ \mbh/$M_{\odot}$}
& \colhead{$\#$ comp.}\\
& & Mpc & mag & mag & mag & & & kpc \\
\colhead{(1)} & \colhead{(2)} & \colhead{(3)}  & \colhead{(4)} & \colhead{(5)} & \colhead{(6)} & \colhead{(7)} & \colhead{(8)} & \colhead{(9)} & \colhead{(10)} & \colhead{(11)}  }
\startdata
3C120        & 0.03301 & 144.9  & 13.49 & 14.62 & 14.62 & 10.41 & 10.41 &  3.26  & 7.74$\pm$0.21 & 2\\
3C390.3      & 0.05610 & 250.5  & 14.90 & 15.76 & 15.76 & 10.45 & 10.45 &  2.48  & 8.46$\pm$0.10 & 2\\
Ark120       & 0.03271 & 143.6  & 13.29 & 14.13 & 15.29 & 10.59 & 10.13 &  0.09  & 8.18$\pm$0.06 & 3\\
Mrk79        & 0.02219 & 96.7	& 14.46 & 14.96 & 15.94 &  9.91 &  9.52 &  0.83  & 7.72$\pm$0.12 & 3\\
Mrk110       & 0.03529 & 155.2  & 15.42 & 16.22 & 17.50 &  9.83 &  9.32 &  0.37  & 7.40$\pm$0.11 & 3\\
Mrk279       & 0.03045 & 133.5  & 14.00 & 14.90 & 16.04 & 10.22 &  9.77 &  0.56  & 7.54$\pm$0.11 & 3\\
Mrk335       & 0.02579 & 112.6  & 14.18 & 15.32 & 16.29 &  9.90 &  9.51 &  0.45  & 7.15$\pm$0.11 & 3\\
Mrk590       & 0.02639 & 115.3  & 14.21 & 14.24 & 15.42 & 10.35 &  9.88 &  1.10  & 7.68$\pm$0.07 & 3\\
Mrk817       & 0.03146 & 138.0  & 14.33 & 14.97 & 17.42 & 10.22 &  9.24 &  0.08  & 7.69$\pm$0.07 & 3\\
PG0052+251   & 0.15500 & 739.2  & 15.02 & 16.25 & 16.25 & 11.24 & 11.24 & 16.76  & 8.57$\pm$0.09 & 2\\
PG0804+761   & 0.10000 & 460.3  & 13.89 & 16.50 & 16.50 & 10.71 & 10.71 &  3.73  & 8.84$\pm$0.05 & 2\\
PG0844+349   & 0.06400 & 287.4  & 14.28 & 16.10 & 16.10 & 10.44 & 10.44 &  3.87  & 7.97$\pm$0.18 & 2\\
PG1211+143   & 0.08090 & 367.6  & 14.43 & 16.93 & 16.93 & 10.33 & 10.33 &  3.06  & 8.16$\pm$0.13 & 2\\
PG1226+023   & 0.15834 & 757.7  & 12.86 & 15.55 & 15.55 & 11.54 & 11.54 &  4.42  & 8.95$\pm$0.09 & 2\\
PG1229+204   & 0.06301 & 282.7  & 15.38 & 15.66 & 16.65 & 10.60 & 10.20 &  1.24  & 7.86$\pm$0.21 & 3\\
PG1411+442   & 0.08960 & 409.5  & 14.58 & 16.80 & 16.80 & 10.48 & 10.48 &  9.52  & 8.65$\pm$0.14 & 2\\
PG1613+658   & 0.12900 & 605.2  & 14.48 & 15.48 & 15.48 & 11.37 & 11.37 & 19.54  & 8.45$\pm$0.20 & 2\\
PG1700+518   & 0.29200 & 1505.1 & 14.87 & 17.84 & 17.84 & 11.41 & 11.41 & 15.79  & 8.89$\pm$0.10 & 2\\
PG2130+099   & 0.06298 & 282.6  & 14.64 & 16.37 & 17.87 & 10.32 &  9.72 &  4.15  & 7.58$\pm$0.17 & 3\\
\enddata
\tablecomments{
Results from imaging of local comparison RM AGN sample.
Details of observations are given in \citet{ben09b}. 
Briefly, all objects considered here were imaged
with HST/ACS, in the F550M filter using the HRC chip.\\
Col. (1): Target ID. 
Col. (2): Redshift.
Col. (3) Luminosity distance in Mpc, based on redshift and the adapted cosmology.
Col. (4): Total extinction-corrected F550M AB magnitude, including PSF (uncertainty of 0.2 mag).
Col. (5): Total host-galaxy extinction-corrected F550M AB magnitude (uncertainty of 0.1 mag). 
Col. (6): Spheroid extinction-corrected F550M AB magnitude (uncertainty of 0.5 mag). 
Col. (7): Logarithm of total host-galaxy luminosity in rest-frame V (solar units), not corrected for evolution.
Col. (8): Logarithm of spheroid luminosity in rest-frame V (solar units), not corrected for evolution.
Col. (9): Spheroid effective radius (in kpc; semi-major axis).
Col. (10): Logarithm of BH mass (solar units) with error, taken from \citet{ben09b}.
Col. (11): Number of components fitted (2=PSF+spheroid; 3=PSF+spheroid+disk).
}
\label{rm}
\end{deluxetable}

\begin{deluxetable}{llccc}
\tabletypesize{\scriptsize}
\tablecolumns{5}
\tablewidth{0pc}
\tablecaption{Fits to the local RM AGN $\log$ \mbh~- $\log$ $L_{\rm sph, V}$ relation}
\tablehead{
\colhead{Method} & \colhead{Sample} & \colhead{$K$} & \colhead{$\alpha$} & \colhead{Scatter}\\
\colhead{(1)} & \colhead{(2)} & \colhead{(3)}  & \colhead{(4)} & \colhead{(5)}}
\startdata
linear fit & this work (with evo.)$^a$                     & -0.07  $\pm$ 0.07 & 0.70 $\pm$ 0.10 & 0.21 $\pm$ 0.08\\
& this work  (no evo.)                                     & -0.11  $\pm$ 0.08 & 0.67 $\pm$ 0.10 & 0.23 $\pm$ 0.09\\
& Bentz et al. (with evo.)                                 &  0.06  $\pm$ 0.06 & 0.72 $\pm$ 0.09 & 0.20 $\pm$ 0.06\\
& Bentz et al. (no evo.)                                   &  0.02  $\pm$ 0.06 & 0.70 $\pm$ 0.08 & 0.20 $\pm$ 0.06\\
BCES & this work (with evo.)                               & -0.12  $\pm$ 0.06 & 0.81 $\pm$ 0.11\\
& this work (no evo.)                                      & -0.15  $\pm$ 0.06 & 0.77 $\pm$ 0.10\\
& Bentz et al. (with evo.)                                 &  0.02  $\pm$ 0.06 & 0.84 $\pm$ 0.09\\
& Bentz et al. (no evo.)$^b$                               & -0.02  $\pm$ 0.06 & 0.80 $\pm$ 0.09\\
\hline
linear fit & this work (host; with evo.)$^a$              & -0.38 $\pm$ 0.12 & 0.96 $\pm$ 0.18 & 0.24 $\pm$ 0.11\\
\enddata
\tablecomments{
Comparison between the different fits (in the form of Equation 5)
to the local RM AGN $\log$ \mbh~- $\log$ $L_{\rm sph, V}$ relation: with and without correction for passive luminosity
evolution, different fitting methods (linear fit vs BCES), and \citet{ben09b} results vs. new analysis in this paper.
In the last row, we give the fit to the local RM AGN $\log$ \mbh~- $\log$ $L_{\rm host, V}$ relation derived in this paper.\\
Col. (1): Fitting method. Linear fit with intrinsic scatter or BCES for comparison with \citet{ben09a}.
Col. (2): Sample. ``evo.'' indicates whether or not data have been corrected for luminosity evolution.
Col. (3): Mean and uncertainty on the best fit intercept.
Col. (4): Mean and uncertainty on the best fit slope.
Col. (5): Mean and uncertainty on the best fit intrinsic scatter (for ``linear fit'' only).\\
$^a$ This is the fit we use in the subsequent analysis.\\
$^b$ This is the fit used in \citet{ben09a}.
}
\label{fits}
\end{deluxetable}

\end{document}